\documentclass[]{aa}
\usepackage[varg]{txfonts}
\usepackage{graphicx,amssymb,amsmath,multirow}
\usepackage[breaklinks,colorlinks,linkcolor=red,citecolor=blue,urlcolor=blue]{hyperref}

\newcommand{\feh}{[Fe/H]}
\newcommand{\afe}{[$\alpha$/Fe]}
\newcommand{\kms}{km s$^{-1}$}
\newcommand{\teff}{$\mathrm{T}_{\mathrm{eff}}$}
\newcommand{\logg}{$\log g$}
\newcommand{\vmic}{$v_{\mathrm{micro}}$}
\newcommand{\vmac}{$v_{\mathrm{mac}}$}
\newcommand{\vsini}{$v{\sin{i}}$}

\newcommand{\micron}{$\mu$m}
\newcommand{\muleo}{$\mu$ Leo}

\newcommand{\FeI}{$\ion{Fe}{i}$}

\begin{document}

\title{An accurate and self-consistent chemical abundance catalogue for the APOGEE/Kepler sample}

\author{
	K. Hawkins\inst{\ref{ioa}}
	\and T. Masseron\inst{\ref{ioa}}
	\and P. Jofr\'e\inst{\ref{ioa},\ref{udp}}
	\and G. Gilmore\inst{\ref{ioa}}
	\and Y. Elsworth\inst{\ref{Birm}}
	\and S. Hekker\inst{\ref{MaxplancGott},\ref{Aarhus}}
	}

\offprints{ \\ 
Keith Hawkins, \email{khawkins@ast.cam.ac.uk}}

\institute{
Institute of Astronomy, University of Cambridge, Madingley Road, Cambridge CB3 0HA, United Kingdom \label{ioa}
\and N\'ucleo de Astronom\'ia, Facultad de Ingenier\'ia, Universidad Diego Portales,  Av. Ej\'ercito 441, Santiago, Chile \label{udp}
\and School of Physics and Astronomy, The University of Birmingham, Birmingham B15 2TT, United Kingdom \label{Birm}
\and Max-Planck-Institut f\"ur Sonnensystemforschung, Justus-von-Liebig-Weg 3, 37077 G\"ottingen, Germany  \label{MaxplancGott}
\and Stellar Astrophysics Centre, Department of Physics and Astronomy, Aarhus University, Ny Munkegade 120, 8000 Aarhus C, Denmark\label{Aarhus} 
}
	
\authorrunning{Hawkins et al.}
\titlerunning{Abundances of APOKASC}
   \date{}

\abstract {The APOGEE survey has obtained high-resolution infrared spectra of more than 100,000 stars. Deriving chemical abundances patterns of these stars is paramount to piecing together the structure of the Milky Way. While the derived chemical abundances have been shown to be precise for most stars, some calibration problems have been reported, in particular for more metal-poor stars. }{In this paper, we aim to (1) re-determine the chemical abundances of the APOGEE+Kepler stellar sample (APOKASC) with an independent procedure, line list and line selection, and high-quality surface gravity information from asteroseismology, and (2) extend the abundance catalogue by including abundances that are not currently reported in the most recent APOGEE release (DR12).}{We fixed the \teff\ and \logg\ to those determined using spectrophotometric and asteroseismic techniques, respectively. We made use of the Brussels Automatic Stellar Parameter (BACCHUS) code to derive the metallicity and broadening parameters for the APOKASC sample. In addition, we derived differential abundances with respect to Arcturus.} {We have validated the BACCHUS code on APOGEE data using several well-known stars, and stars from open and globular clusters. We also provide the abundances of C, N, O, Mg, Ca, Si, Ti, S, Al, Na, Ni, Mn, Fe, K, and V for every star and line, and show the impact of line selection on the final abundances. Improvements have been made for some elements (e.g. Ti, Si, V). Additionally, we measure new abundance ratios not found in the current APOGEE release including P, Cu, Rb, and Yb, which are only upper limits at this time, as well as Co and Cr which are promising.}{In this paper, we present an independent analysis of the APOKASC sample and provide abundances of up to 21 elements. This catalogue can be used not only to study chemical abundance patterns of the Galaxy but also to train data driven spectral approaches which can improve the abundance precision in a restricted dataset, but also full APOGEE sample. } 
\maketitle

\section{Introduction}
The Milky Way is a complex system and is known to host several structural components. Over the last few decades, it has been shown, with small-to-modest samples of local stars, that some of these components may be chemically distinct from one another \citep[e.g.][]{Edvardsson1993, Fuhrmann1998, Venn2004, Nissen2010, Sheffield2012, Ramirez2012, Feltzing2013, Bensby2014}. The advent of large multi-object spectroscopic surveys, such as the Apache Point Observatory Galactic Evolution Experiment \citep[henceforth APOGEE,][]{Eisenstein2011, Majewski2015}, the Gaia-ESO survey \citep[henceforth GES,][]{Gilmore2012}, the Australian GALAH survey \citep{De_silva2015}, and others, have greatly supplemented and advanced these local samples. In particular, these aforementioned surveys are collecting large samples ($\sim10^5$) of high-resolution (R~=~$\lambda/\Delta\lambda \sim$~20,000 -- 60,000) spectra which, with the help of automatic stellar parameters and abundance pipelines, have enabled homogenous bulk Galactic chemical evolution studies \citep[e.g.][]{Nidever2014, Recio-Blanco2014, Mikolaitis2014, Masseron2015, Hawkins2015b}. 

The spectra from these surveys have been used to homogeneously derive the basic stellar parameters, effective temperature (\teff), surface gravity (\logg), metallicity (\feh), and, in some cases, microturbulent velocity (\vmic) for up to 100,000 stars. In addition, these surveys have produced up to 34 elemental abundances for a sizable fraction of the sampled stars. These parameters, and the chemical abundances in particular, are integral to study the nature and structure of our Galaxy as they provide useful `tags' as to the environment the stars were born in \citep[e.g.][]{Freeman2002, Hawkins2015b, Hogg2016}.

In the context of chemical abundance patterns in the Milky Way, the SDSS-III/APOGEE project \citep{Majewski2015} has been transformative because it not only surveys a large volume within the Galaxy, thanks to its targeting of giant stars, but it also has done high-resolution $H$-band spectroscopy and delivered stellar parameters and chemical abundance of up to 15 elements. This has made it possible to study the interfaces of Galactic components \citep[e.g.][]{Nidever2014, Hayden2015, Hawkins2015b, Masseron2015} and their ages \citep{Martig2016}. In particular, several interesting conclusions have been made using the APOGEE data including the confirmation of the existence of a chemically distinct accreted halo \citep[e.g.][]{Nissen2010, Hawkins2014}, evidence of a metal-poor thin disk \citep{Hawkins2015b}, differing star formation rates in the thin and thick disks across all metallicities \citep{Masseron2015}, a positively skewed metallicity distribution function in the outer galaxy indicating the importance of radial migration \citep{Hayden2015}, and chemical tagging of phase-space substructures \citep{Hogg2016}. These conclusions require chemical abundances, and in particular the metallicity, to be at least precise if not accurate. However, in addition to there being known issues with the APOGEE DR12 \feh\ calibration worsening at lower metallicities, there are also large systematic zero-point offsets in some elements with respect to literature \citep{Holtzman2015}.

Thus, the primary aim of this work is to solve the metallicity calibration issue seen in APOGEE DR12 down to metallicities around --1.0 dex for a subsample of the data. We present in this paper an independent analysis using APOGEE spectra of Kepler targets (henceforth, the APOKASC sample) with several improvements in the atomic and molecular input data, elemental line selection, and a line-by-line differential analysis. In addition, the \logg\ for the subsample of interest are determined independent of spectroscopy, via asteroseismology \citep{Hekker2013}, and the \teff\ is derived using the spectra but corrected using photometry.  We used these to determine the remaining stellar parameters (\feh, \vmic) and chemical abundances of up to 20 elements including elements, such as Co, which are currently not provided by the APOGEE stellar parameter and chemical abundance pipeline (ASPCAP). The improved abundance ratios and determination of broadening parameters are crucial to extract further information about Galactic evolution from the APOGEE survey.  

We organize this paper in the following way: In Sect. \ref{sec:data}, we describe the APOGEE spectral data for the APOKASC sample and the Brussels Automatic Code for Characterising High accUracy Spectra (BACCHUS) pipeline which is used to derive the metallicity, broadening parameters, and chemical abundances. In that section, we also discuss the validation of the pipeline using a sample of Gaia benchmark stars and open and globular clusters. In Sect. \ref{sec:results} we present the stellar parameters and chemical abundance for up to 21 elements for the APOKASC sample. We then discuss these results in the context of the literature and the APOGEE survey in Sect. \ref{sec:discussion}. Finally we summarize, conclude, and discuss future extensions to this project in Sect. \ref{sec:conclusion}.

\section{Data and Method} 
In Sect. \ref{subsec:spectraldata} we introduce the APOGEE survey and the properties of the spectral data. We then describe, in Sect. \ref{subsec:BACCHUS}, the BACCHUS pipeline which was used to derive the broadening parameters, metallicity, and chemical abundances. The validation of the pipeline using both benchmark stars and globular and open clusters is described in Sect. \ref{subsec:Validation}. 
\label{sec:data}
\subsection{Spectral Data}
\label{subsec:spectraldata}
\begin{figure}
	 \includegraphics[width=0.83\columnwidth,angle =270]{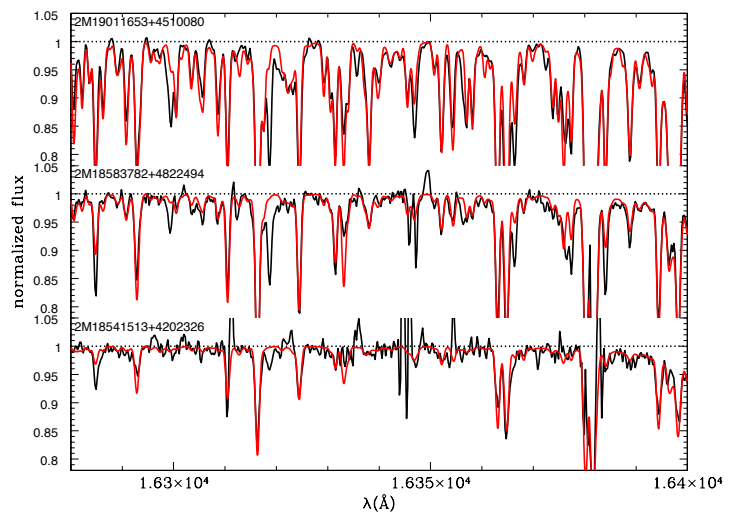}
	\caption{The continuum normalized observed spectra (black) and best fit synthetic spectra (red) for three stars with different stellar parameters and SNR including 2M19011653+4510080, 2M18583782+4822494, and 2M18541513+4202326 from top to bottom, respectively. 2M19011653+4510080 has \teff\ = 4544~K, \logg\ = 2.54~dex, \feh\ = +0.53~dex, SNR = 874~pixel$^{-1}$, 2M18583782+4822494 has \teff\ = 4740~K, \logg\ = 2.54~dex, \feh\ = --0.26~dex, SNR = 225~pixel$^{-1}$, and 2M18541513+4202326 has \teff\ = 4969~K, \logg\ = 2.44~dex, \feh\ = --0.82~dex, SNR = 134~pixel$^{-1}$.  }
	\label{fig:normspec}
\end{figure}

We have made use of a Kepler subsample of the twelfth data release (DR12) of the SDSS III-APOGEE survey \citep[details of the APOGEE survey can be found in][]{Eisenstein2011,Majewski2015}. This subsample contains nearly 2000 stars. The APOGEE survey has collected a large number ($\sim$10$^5$) of high-resolution (R$\sim$ 22,500) spectra. The spectra used for the APOGEE survey were taken using a fiber-fed infrared spectrograph which covers the $H$-band between 1.51 and 1.70 \micron. The publicly available combined spectra\footnote{Details on the combined spectra, which are found in the DR12 `apStar' data product, and how to obtain the data can be found at \url{http://www.sdss.org/dr12/irspec/spectral_combination/} and \url{http://www.sdss.org/dr12/irspec/spectro_data/}, respectively.} were used in this study. The spectra are characterized by a signal-to-noise ratio (SNR) which ranges from 70 to more than 800 pixel$^{-1}$. The typical SNR is around 200 pixel$^{-1}$. These spectra are the product of a combination of all of the visits that APOGEE has made to each star. The spectra have been radial velocity (RV) corrected and resampled to common wavelength sampling before being combined. A weighted combination of all of the spectra is then computed and is used in this work. These combined spectra have not been continuum normalized. That is accomplished by using the BACCHUS pipeline. The continuum normalization is done locally by fitting a linear function to a selected set of continuum points over a 30 \AA\ window. Continuum points are selected automatically using the spectral synthesis as a guide. An example of the continuum normalized spectra using the BACCHUS pipeline for three stars, which have different stellar parameters and SNR, can be found in Fig. \ref{fig:normspec}.  

The survey team has released stellar parameters (\teff, \logg\ and \feh) and abundances for up to 15 chemical species \citep[][]{Holtzman2015, GarciaPerez2015}. We refer the reader to the ASPCAP description paper by \cite{GarciaPerez2015} and its implementation for DR12 \cite{Holtzman2015} for more information. The stellar parameters have been derived by interpolating within a grid of synthetic spectra \citep[see section 4.1 of][]{Holtzman2015} using the FERRE code \citep{AllendePrieto2006}. However, the \vmic\ are fixed to a linear relationship with \logg. This relationship was derived using a subset of stars within the APOGEE survey. Figure 2 of \cite{Holtzman2015} suggests that there is likely also a metallicity dependence as well. This may be a possible explanation, at least in part, for the metallicity calibration that is required to the ASPCAP values. To test this, we have solved the broadening parameters in this work.

\subsection{The BACCHUS code}
\label{subsec:BACCHUS}
The BACCHUS code \citep{Masseron2016} consists of three different modules that are designed to derive equivalent widths, stellar parameters, and abundances. Because we wanted to take full advantage of the asteroseismic data, we have fixed \logg\ throughout the analysis process to those determined by \cite{Pinsonneault2014}. The \teff\ is also fixed and is selected to corrected ASPCAP \teff\ from DR10 to be consistent with the asterosismic \logg\ values. As a reminder, the \teff\ were corrected by comparing the values determined from the ASPCAP pipeline and the value computed using the 2MASS ($J-K_s$)-\teff\ relationship from \cite{Gonzalez-Hernandez2009}. The \logg\ were taken from the asteroseismic scaling relations using the selected \teff\ \citep[see Sect.4 and Equation 2 of][for more details]{Pinsonneault2014}. The approach of fixing \teff\ and \logg\ to values determined independently of spectroscopy and deriving chemical abundances has been successfully applied to optical spectra with the BACCHUS code on the set of reference stars called Gaia benchmark stars, which are key calibrators of the Gaia-ESO survey \citep[e.g.][]{Jofre2014, Jofre2015, Hawkins2016a}.

 The current version of the BACCHUS code relies on the radiative transfer code Turbospectrum \citep{Alvarez1998,Plez2012} and the MARCS model atmosphere grid \citep{Gustafsson2008}. One particular asset of Turbospectrum is its ability to handle radiative transfer in spherical geometry, recommended when dealing with giants. 
With fixed \teff\ and \logg, the first step consists of determining the metallicity, the \vmic\ parameter, and the convolution parameter. The metallicity provided is the average abundance of selected Fe lines. The \vmic\ is obtained by minimising the trend of Fe abundances against their reduced equivalent width (REW). The convolution parameter stands for the total effect of the instrument resolution, the macroturbulence, and \vsini\ on the line broadening. However, given the quality of the data, we could not disentangle each of those effects. Furthermore, we note that the instrument resolution varies as a function of fiber position \citep{Holtzman2015}. Therefore, we derive one single global convolution value per spectrum, based on the average broadening of Fe lines. For this, we assume a gaussian convolution profile. Once metallicity, microturbulence, and convolution parameters are determined, O, C, and N are derived. Indeed, the line opacities of those elements dominate the APOGEE spectra via the CO, OH, and CN molecules and thus must be taken into account when fitting any part of the spectrum. Once those elements are measured, the whole process is iterated until convergence.    

For each element and each line, the abundance determination module then proceeds in the following way: (i) a spectrum synthesis, using the full set of (atomic and molecular) lines, is used to find the local continuum level via a linear fit, (ii) cosmic and telluric rejections are performed, (iii) the local S/N is estimated, (iv) a series of flux points contributing to a given absorption line is automatically selected, and (v) abundances are then derived by comparing the observed spectrum with a set of convolved synthetic spectra characterised by different abundances. Four different abundance determinations are used: (i) line-profile fitting, (ii) core line intensity comparison, (iii) global goodness-of-fit estimate (aka $\chi^2$), and (iv) equivalent width comparison. Each diagnostic yields validation flags. Based on these flags, a decision tree then rejects the line or accepts it, keeping the best-fit abundance. We adopted the $\chi^2$ diagnostic as the abundance because, by experience, it is the most robust. However, we store the information from the other diagnostics, including the standard deviation between all fourth methods (which we refer to as the method-to-method scatter), in order to aid in the line selection (see Sect. \ref{subsec:lineselection}). \\

\subsection{Linelists} \label{subsec:linelist}
The linelists employed for the synthesis are the following: for atoms, the most recent release of VALD data \citep{VALD2015} has been used has a basis. Hyperfine structure has been added for Co \citep{Pickering1996}, V \citep{Unkel1989,Palmeri1995,Palmeri1997}, and Mn \citep{Blackwell2005} as well as isotopic shift information for Cu \citep{Elbel1961,Bergstroem1989,Bengtsson1990}. Concerning molecular linelists, we include OH \citep{Brooke2016}, CN \citep{Sneden2014}, CO \citep{Rothman2010}, and their respective carbon isotopologs, as well as MgH \citep{Yadin2012}, NH \citep{Brooke2016}, CH \citep{Masseron2014}, C$_2$ (P. Quercy, private communication), SiO \citep{Barton2013}, and CaH \citep{Yadin2012}. Note that, in contrast to \citet{Shetrone2015}, we chose not to apply any empirical correction on line position, $\log gf$, or collisional broadening parameters for the linelist used in this study. As detailed in further sections, we rather make a careful line selection as well as providing abundances based on a line-by-line differential approach. 

\subsection{Line Selection}
\label{subsec:lineselection}
Although the pipeline has its own procedure to include or reject lines on a star-by-star basis, it is still important to select the lines beforehand because of the uncertainty related to the synthesis approach such as strong NLTE and/or 3D effects as well as line saturation. 
The initial line selection for iron was done by searching for all Fe lines with theoretical (i.e. synthetic) equivalent widths (EW) larger than 5 m\AA. These lines were then synthesized for the Sun and Arcturus using BACCHUS assuming the solar parameters \teff\ = 5777 K, \logg\ = 4.44 dex, \feh\ = 0.00 dex, and \vmic\ = 0.86 \kms\ and Arcturus stellar parameters of \teff\ = 4286 K, \logg\ = 1.64 dex, \feh\ = --0.52 dex, and \vmic\ = 1.6 \kms. We note that we assume the solar chemical composition of \cite{Asplund2005}. We then compared these synthesized Fe lines to a high-resolution infrared (R $\sim$ 100,000) Arcturus atlas \citep{Hinkle2005} and Solar atlas \citep{Hinkle1995}. The selected Fe lines were visually inspected to ensure that the spectral fit was adequate. 

In addition, lines were rejected if they were found to have Fe abundances outside of $\pm$0.10~dex of the solar value (log($\epsilon_{\mathrm{Fe}}$) = 7.45). This was done to avoid selecting Fe lines where the fit was not good, or the atomic data was not adequate. In total, there are 20 \FeI\ lines that were selected for the determination of \feh\ and \vmic, which was derived by forcing the correlation between the REW and \feh\ to be zero. Many of these lines were studied in the work of \cite{Smith2013}, however, we included additional lines. The abundance of all selected Fe lines for every star in the APOKASC sample can be found in the provided online tables (see Appendix~\ref{app:A} for more details).

\begin{figure}
	 \includegraphics[width=\columnwidth]{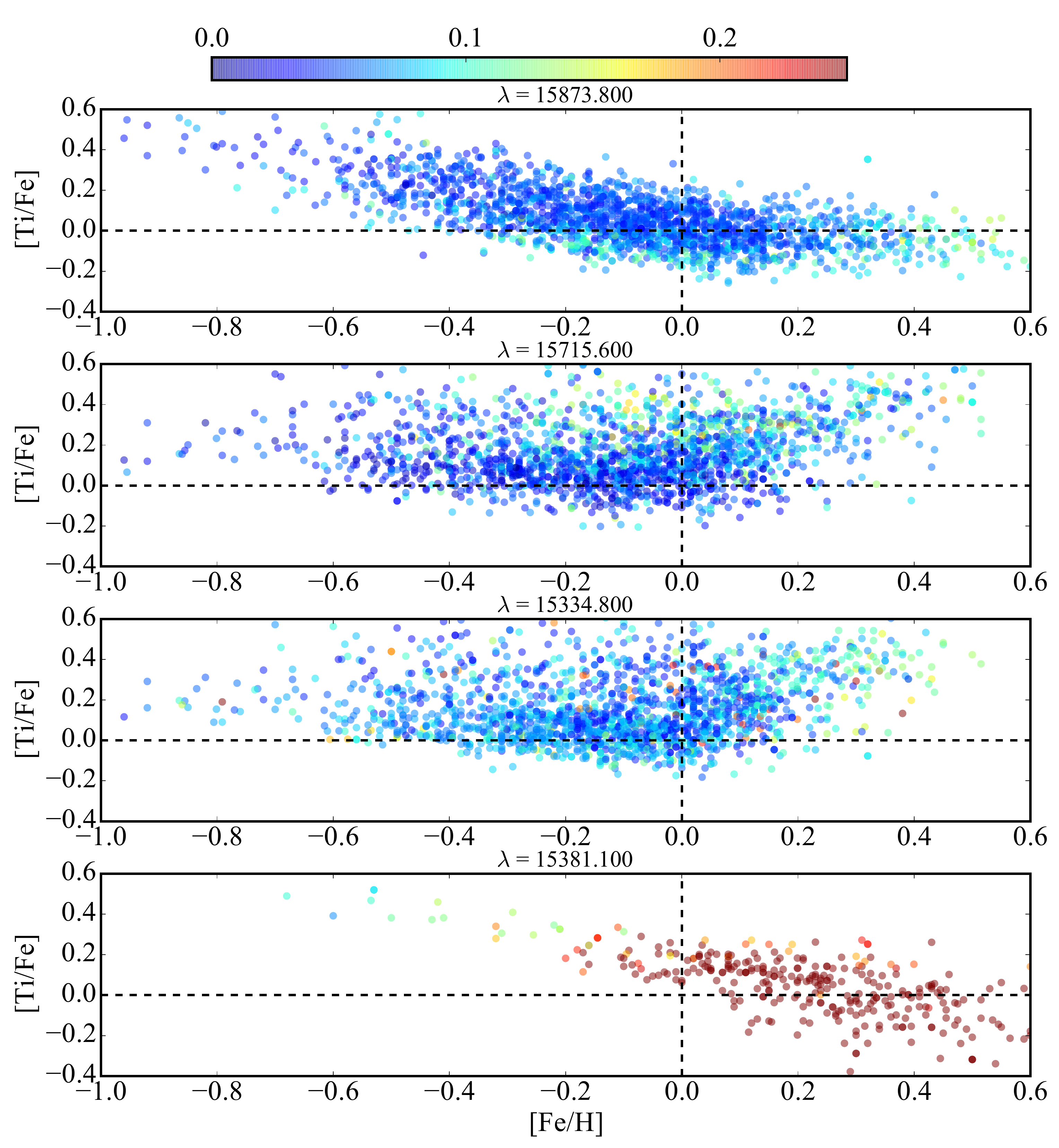}
	\caption{The [Ti/Fe] as a function of metallicity for the full APOKASC sample for the 15873.8~\AA, 15715.6~\AA, 15334.8~\AA, and 15381.1~\AA\ lines from top to bottom, respectively. The stars are color-coded by the BACCHUS method-to-method dispersion, which is defined as the standard deviation of the abundance derived from the four procedures.}
	\label{fig:tilines}
\end{figure}

The line selection for the other elements was done in a similar way as above. Lines for each element with theoretical equivalent widths (EW) larger than 5 m\AA\ in a synthetic Arcturus spectrum were initially selected. Additionally, each line was also measured in every star in the APOKASC sample whether it was chosen for the final selection or not.  Lines were first visually inspected for a good fit in the Sun and Arcturus. If the line was not well reproduced by the synthesis it was rejected. In addition, for most elements, lines were also rejected if they were flagged as problematic by the BACCHUS pipeline in a substantial fraction of the APOKASC sample. Finally, lines that were very discrepant to other selected lines were discarded. 

In Fig. \ref{fig:tilines}, we present an example of a unique diagnostic diagram that has been used to aid the line selection. In the figure, we plot the [Ti/Fe] as a function of metallicity for 4 Ti lines (15873.8~\AA, 15715.6~\AA, 15334.8~\AA, and 15381.1~\AA\ lines from top to bottom, respectively) of the 31 lines initially selected for every star in the sample. Each circle represents a star and it is color-coded by the method-to-method scatter (described in Sect \ref{subsec:BACCHUS}). Circles that are colored blue have low method-to-method scatter (i.e. all four procedures to measure abundance in BACCHUS agree well) and circles that are colored red have high method-to-method scatter (i.e. the methods disagree indicating the line fit may not be of good quality). 

This diagram has been used to make careful line selection choices. For example, we have deselected the Ti line at 15381.1~\AA\ for two reasons: (1) there is only a small subset (about 15\%) of the data where this line is not flagged as a poor fit in the BACCHUS decision tree, and (2) in this small subset the method-to-method scatter is very large (as indicated by the red colored circles). Interestingly, the three lines that are left give vastly different [Ti/Fe]-[Fe/H] trends from each other. From the literature \citep[e.g.][]{Bensby2014} we know that [Ti/Fe] increases with decreasing metallicity at sub-solar metallicities and is roughly flat at super-solar metallicities. This is only the case for the Ti line at 15873.8~\AA\ (top panel of Fig \ref{fig:tilines}). So why are the other two lines (15715.6~\AA\ and 15334.8~\AA) so discrepant? This is likely due to NLTE or saturation effects as both lines are very strong. As such, these lines were rejected. This powerful diagnostic plot was constructed for every element to study the affect that different line selections would have on the chemical abundance patterns observed and was also used to aid the line selection process. 

\begin{figure}
	 \includegraphics[width=\columnwidth]{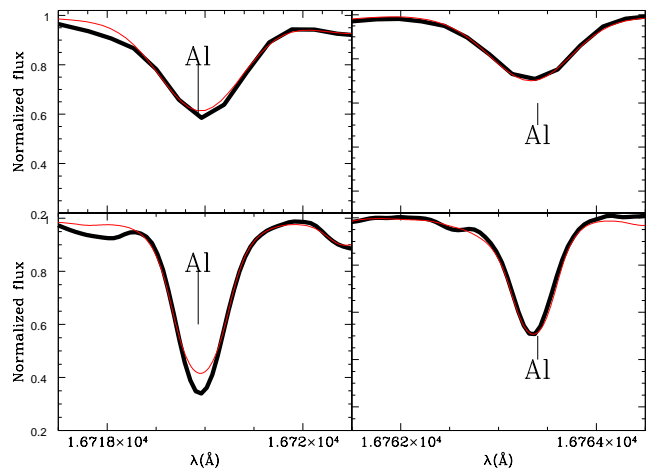}
	\caption{Two Al lines in Arcturus high resolution spectrum (black thick) against synthesis (red thin). The upper panel shows the lines at the APOGEE resolution while bottom panel shows the line at high-resolution.}
	\label{fig:Al_lines}
\end{figure}

In addition to those diagnosis plots, we also systematically synthesised every element and line in Arcturus at high-resolution to compare to with the Hinkle atlas. As an example, we show in Fig.~\ref{fig:Al_lines} the example of the two Al lines as used in the DR12 APOGEE release. It is striking that while the line  at 16763 \AA\ is well reproduced at high resolution, the 16719~\AA\ line is poorly fit in the core. This is a strong indicator of NLTE effects occurring for that line. Therefore, despite the apparent good quality of the fit in the APOGEE spectrum, we reject the 16719~\AA\ line and use only the 16763~\AA\ line in this study. Beyond Ti and Al, the other element where we had strong indication of 3D and/or NLTE effects is S. Thus, we rejected the line at 15469.8~\AA\ and selected only 15478.5~\AA. 

For all elements, the final line selection had, on average, between one and a five lines per element. Unlike for Fe, we did not ensure that the derived solar abundance of each element in each line were within 0.10 dex of the solar value. To get around this, we implemented a line-by-line differential analysis, with respect to Arcturus, in order to improve precision \citep[e.g. see][for an extensive discussion on how differential analyses can improve precision]{Jofre2015}. The abundances per line for every star can be found in the online tables.


\subsection{Validation}
\label{subsec:Validation}
With the lines selection in place, we proceed to validate the procedure. We have done two sets of validation tests to quantify the performance of the BACCHUS pipeline, in particular on APOGEE data. The first test was done using the three {\it Gaia} benchmark stars the Sun, Arcturus, and \muleo. These are well-studied stars which have \teff\ and \logg\ measured independently from spectroscopy. For more information on the absolute stellar parameters of the benchmark stars and the procedures used to determine them we refer the reader to the {\it Gaia} benchmark papers \citep[][]{Heiter2015, Jofre2014, Jofre2015, Hawkins2016a}. This test is outlined in Sect. \ref{subsubsec:Benchmark}

The second test made use of a sample of 119 stars in eight globular and open clusters. We compared the mean metallicity derived from BACCHUS with the literature values to infer its performance. This test is described in Sect. \ref{subsubsec:GC}
\subsubsection{Benchmark Stars: The Sun, Arcturus, and \muleo}
\label{subsubsec:Benchmark}
One way to validate the pipeline is to determine how well it retrieves the stellar parameters of a set of well-known or benchmark stars. For this test we use three benchmark stars defined in a series of papers on the {\it Gaia} Benchmark star project \citep[][]{Heiter2015, Jofre2014, Jofre2015, Hawkins2016a}. The work of \cite{Heiter2015} discusses the \teff\ and \logg\ determination of the benchmark stars. The literature metallicity of the benchmark stars are sourced from \cite{Jofre2014}. 

The three benchmark stars that were chosen were the Sun, Arcturus, and \muleo\ as these are the only benchmarks which have APOGEE spectra public. The Sun was chosen because it is our nearest star and the one with the highest quality parameters. However the Sun is a dwarf while the stars in the APOKASC sample are giants. Arcturus was chosen because it represents a red-giant star that was suggested by \cite{Jofre2015} for differential analysis. Finally, \muleo\ was chosen because it is a metal-rich red-giant star. The other giant benchmark stars were not observed in DR12.  The results from the benchmark validation analysis are summarized in Table \ref{tab:benchmark}. 
 \begin{table}
 \setlength{\tabcolsep}{4pt}

\caption{Benchmark Stars Stellar Parameters.} 
\begin{tabular}{l l l l l l l} 
\hline\hline
Star & \teff$^a$ & \logg$^a$ & \feh\ & \vmic & \feh$_{\mathrm{lit}}^{b}$& \vmic$_{\mathrm{lit}}^{b}$\\
 & (K) & (dex) & (dex) & (\kms) & (dex) & (\kms)\\
\hline\hline 
Sun & 5777 & 4.44 & --0.01 & 0.76 &+0.03 & 1.2\\
Arcturus & 4286 & 1.64 & --0.54 & 1.21&--0.52& 1.3\\
\muleo & 4474 & 2.51 & +0.27 & 1.07&+0.25 & 1.1\\
\hline \hline
\end{tabular}
\\ \\
\tablefoot{($^a$) denotes parameters that have been taken from \cite{Heiter2015}. ($^b$) denoted the \feh\ and \vmic\ are taken from \cite{Jofre2014}. }
\label{tab:benchmark}
\end{table}

We fixed \teff\ and \logg\ of the Sun to 5777~K and 4.44~dex, respectively and derived \vmic\ and \feh\ using the selected Fe lines (for a description of the line selection consult Sect. \ref{subsec:lineselection}) in order to test the validity of not only the line selection but also the BACCHUS procedure. Using these parameters we recovered a solar metallically of \feh\ = --0.01 dex $\pm$ 0.08 dex and a \vmic\ of 0.76 $\pm$ 0.07 \kms\ consistent with the literature \citep[e.g.][]{Jofre2014}.  

The \teff\ and \logg\ of Arcturus were set to 4286~K and 1.64~dex, respectively and the \vmic\ and \feh\ were derived. Using these parameters we recovered a metallically for Arcturus of \feh\ = --0.54 dex $\pm$ 0.09 dex and a \vmic\ of 1.21 $\pm$ 0.10 \kms. These values are in good agreement with the literature \citep[e.g.][]{Jofre2014}. 

In addition, we choose Arcturus as the reference star for the differential chemical abundance analysis that we implemented in Sect. \ref{subsec:Chemistry}. This was done to improve the precision in the chemical abundances by effectively correcting systematics induced by the inaccuracies in the line list. The abundances that are derived for Arcturus using BACCHUS, found in Table \ref{tab:Arcturuschem}, are generally in good agreement with the literature. There are some cases, Mn for example, which have up to a 0.20 dex offset in the abundance determined from BACCHUS and the literature. This may result from inaccurate optical and/or infrared line lists or hyperfine structure effects. 

The benchmark star \muleo\ supplements Arcturus because of its relatively high metallicity \citep[\feh\ = +0.25 $\pm$ 0.15 dex][]{Jofre2014}. We fixed the \teff\ to 4474~K and the \logg\ to 2.51~dex \citep[e.g.][]{Heiter2015}. Using these parameters we performed the same procedure used for the APOKASC sample. The pipeline yielded a \feh\ = +0.27 dex and a \vmic\ = 1.07 $\pm$ 0.10 \kms. These values are consistent with the literature on this star \citep[e.g.][]{Jofre2014}. 

\begin{figure}
	 \includegraphics[width=\columnwidth]{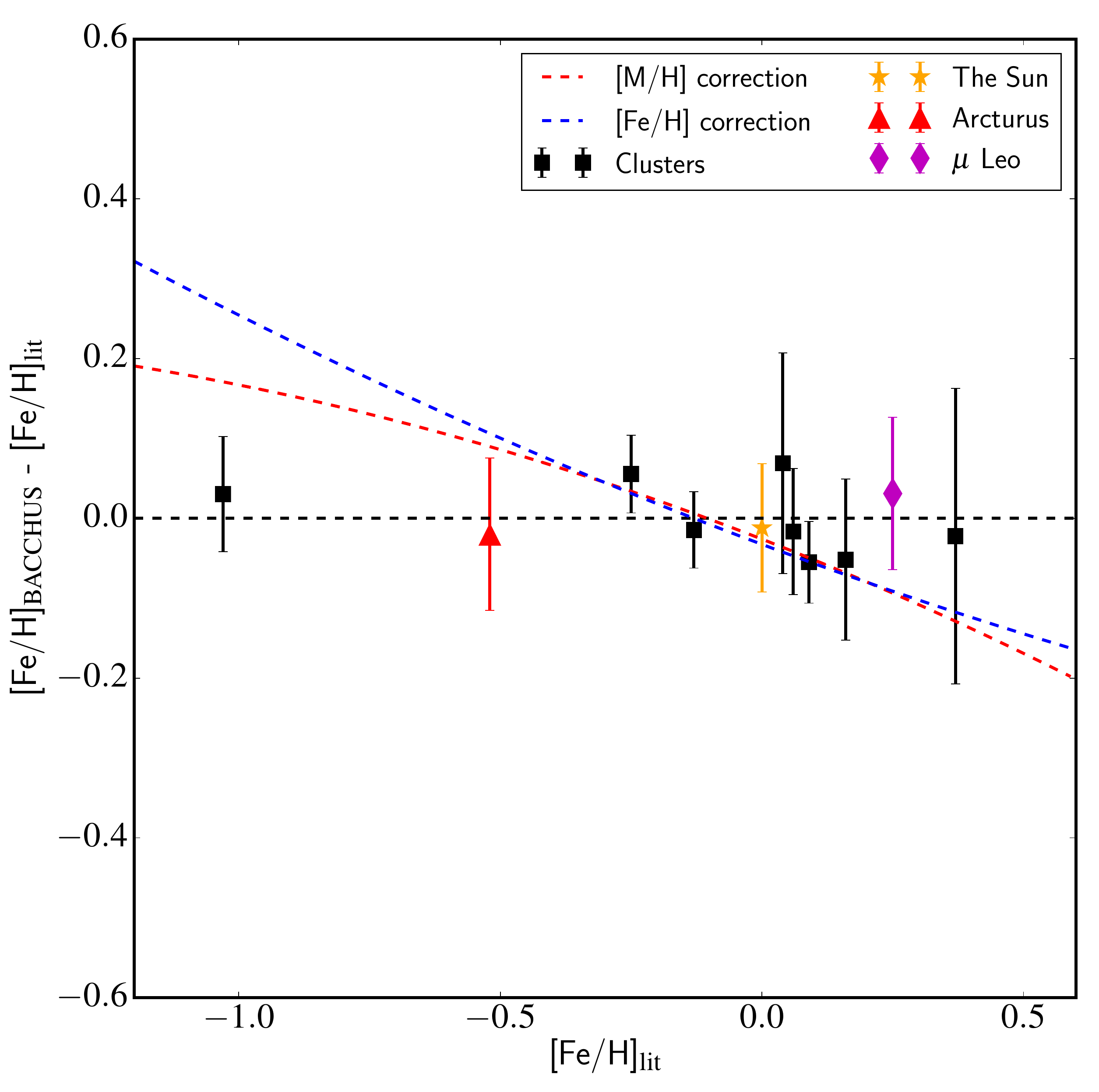}
	\caption{The difference between the \feh\ derived of the eight clusters \citep[NGC6791, NGC6819, NGC7789, M67, NGC188, NGC2420, NGC2158, and M107,][]{Meszaros2013}, derived in this study with the reference value (see Table \ref{tab:clusters}) as a function of the literature \feh. In addition, the Sun and Arcturus are shown as an orange star and red triangle, respectively. The blue and red dotted lines represent the [M/H] and \feh\ corrections, respectively, that are derived in \cite{Holtzman2015} for comparison.}
	\label{fig:clusters}
\end{figure}

Fig.~\ref{fig:clusters} shows the difference of the derived metallicity from the BACCHUS pipeline and the literature metallicity as a function of the literature metallicity for the Sun, which shown as an orange star, Arcturus, which is shown as a red triangle, and \muleo, which is shown as a magenta diamond. This figure indicates that the pipeline recovers the metallicity of these benchmark stars within 0.04 dex of their literature values. 

\subsubsection{Open and Globular Clusters}
\label{subsubsec:GC}
Globular and open clusters offer a great opportunity to quantify the validity of the BACCHUS pipeline which we use to derive the metallicity, chemical abundances, and broadening parameters for the APOKASC sample. For the sake of comparison with the DR12 of APOGEE, we have analyzed the spectra of 119 stars in eight globular and open clusters spanning the metallicity range +0.4 $<$ \feh\ $<$ --1.05. These clusters and their members were selected from \cite{Meszaros2013}. We refer the reader to Sects. 2 and 3 of \cite{Meszaros2013} for a detailed discussion on cluster members, observations, data reduction, and the analysis of these stars with ASPCAP.

Since we fixed the \teff\ and \logg\ of the stars in the APOKASC sample, we also do the same for the cluster stars. In this way, we ensure that the pipeline treats these cluster stars in the same way as the APOKASC sample. Most of the cluster stars do not have seismic \logg\ estimates and thus a caveat to this analysis is that we must assume the \teff\ and \logg\ of APOGEE DR12 which, at least for \logg, is derived differently from the APOKASC sample. There are 28 stars in two metal-rich clusters (NGC6791 and NGC6819) which have publicly available seismic \logg\ information in the APOKASC catalog. The mean metallicity of these two clusters are consistent when using both the seismic or corrected ASPCAP\footnote{We remind the reader that the ASPCAP \logg\ values have been corrected using the APOKASC seismic information \citep[see Sect. 5.2 of][]{Holtzman2015}. }.  

Table \ref{tab:clusters} contains the mean metallicity from the literature of the eight clusters that have been analyzed \citep[e.g.][]{Harris1996, Bragaglia2001, Carraro2006,Jacobson2011b}. It also contains the mean metallicity we derive from BACCHUS for a certain number of stars within each cluster. It is important to keep in mind that while clusters are often cited as a great way to calibrate spectral pipelines, some clusters can have widely varied mean metallicities within the literature. For example, the work of \cite{Heiter2014} explored the status of the mean metallicity of many open clusters through the literature. They found that for at least three of the clusters analyzed in both APOGEE and this work, including NGC188, NGC2158, and NGC2420, the mean metallicity within the literature vary by as much as 0.20 dex between medium- and high-resolution studies. Specifically, NGC2158 has a metallicity of \feh\ = --0.03 in \cite{Jacobson2009} based on one star, yet a significantly lower metallicity of \feh\ = --0.28~dex in \cite{Jacobson2011b} based on 15 stars. The discrepancy is attributed the choice of reddening E($B-V$) which in turn changes the \teff\ and \feh\ of the stars in their sample. We choose to adopt the value from \cite{Jacobson2011b} in part because it is based on a larger number of stars. 

In addition to the benchmark stars, in Fig.~\ref{fig:clusters}, we display the difference of the mean cluster metallicities found in this work and the literature value. We also plot the correction formulae that are derived for the [M/H] and [Fe/H] parameters based on these same clusters in \cite{Holtzman2015}. This figure indicates that we adequately recover the mean literature metallicity of the eight clusters within the typical uncertainty of \feh. Fig.~\ref{fig:clusters} indicates that we do not need to apply a metallicity correction formula to force agreement with literature for metal-poor stars. The improvements we have implemented, namely a careful line selection and deriving the broadening parameters, might explain why APOGEE requires such a correction formula. In Sect \ref{subsec:clusterchem}, we derived the abundances of up to 21 elements for each cluster and use those results to provide additional tests of the abundance precision.

 \begin{table}
\caption{Calibration Clusters.} 
\begin{tabular}{l c c c c c} 
\hline\hline 
Cluster & \feh$_{\mathrm{Lit}}$&\feh&$\sigma$\feh&N&Reference\\
 & dex & dex & dex & \\
 \hline\hline \\
NGC6791&+0.37&+0.35&0.15&23&(1)\\
NGC6819&+0.16&+0.11&0.10&28&(2)\\
NGC7789&+0.09&+0.03&0.05&5&(3)\\
M67&+0.06&+0.04&0.08&23&(3)\\
NGC188&+0.04&+0.11&0.14&5&(3)\\
NGC2420&--0.13&--0.14&0.05&9&(3)\\
NGC2158&--0.24&--0.19&0.05&10&(3)\\
M107&--1.03&--1.01&0.07&16&(4)\\
 \hline

\hline \hline
\end{tabular}
\\ \\
\tablefoot{The literature mean metallicity, \feh$_{\mathrm{Lit}}$, are sourced from the following: (1) \cite{Carraro2006}, (2) \cite{Bragaglia2001}, (3) \cite{Jacobson2011b}, and (4) 2010 version of \cite{Harris1996}. The \feh\ and $\sigma$\feh\ are the mean and star-to-star dispersion of each cluster, respectively.}
\label{tab:clusters}
\end{table}

\subsection{Differential Analysis} \label{subsec:differential}
Unlike most large surveys, including APOGEE, we have implemented a line-by-line differential analysis with respect to a reference star. This procedure has been shown to be a way to improve abundance precision (and possibly accuracy) by accounting for systematics caused by inaccuracies in the line list and other effects \cite[e.g.][]{Bensby2014, Ramierez2014, Jofre2015}. Under a differential approach we compare, line-by-line, the abundance of each star with that of the reference star. This leaves the derived abundances for each star relative to the reference star. To convert back to a solar normalised value we must assume an [X/H] value for the reference star. However, we note that the [X/H] value is just a zero-point scaling factor but does not affect the overall abundance precision. 

Arcturus is used as the reference star for differential analysis because the APOKASC sample are red-giant stars with metallicities around --0.10 dex not to dissimilar to Arcturus. In addition, it has measured chemical abundances from various sources in both the optical and infrared regimes. We derived the chemical abundances for a total of 21 chemical species.

Additionally, we have shown in Sect.~\ref{subsec:Validation} that \feh\ values for the benchmark stars and clusters derived are well reproduced as a result of selecting lines which have good log gf values. We therefore do not apply a differential analysis on Fe. For the remaining  elements, including C, N, O, Mg, Ca, Si, Ti, S, Al, Ni, Na, Mn, K, Cr, Co, Cu, and V, we have implemented a line-by-line differential analysis whereby we compare directly the abundances of each star with the values derived in Arcturus. The abundances are then solar-scaled using the adopted [X/H] abundances of these elements for Arcturus shown in the top part of Table \ref{tab:Arcturuschem}. We note that we could not apply a differential analysis to P because there are very few phosphorus measurements in literature. Therefore, we have no independent way of checking that the absolute scale in our [P/Fe] diagrams are valid. 

\begin{table*}
\caption{Arcturus Chemical Abundances.} 
\begin{tabular}{c c c c c c c c c c} 
\hline\hline
Element&[X/H]&$\sigma$&N&Jofr\'e$^{a}$&Holtzman$^{b}$&Smith$^{c}$&Ram\'irez$^{d}$&[X/Fe]& Adopted [X/H]\\
\hline\hline
 & & & &  &Differential & & & \\
 \hline
 C&--0.41&0.08&4&...&--0.43&--0.43&--0.09&0.10&--0.41\\
N&--0.10&0.09&16&...&--0.52&--0.14&...&0.41&--0.14\\
O&--0.14&0.03&8&...&--0.32&--0.02&--0.02&0.37&--0.02\\
Mg&--0.09&0.06&6&--0.15&--0.40&--0.38&--0.15&0.42&--0.15\\
Ca&--0.44&0.05&3&--0.41&--0.51&--0.47&--0.41&0.07&--0.41\\
Si&--0.17&0.04&5&--0.25&--0.33&--0.39&--0.19&0.34&--0.25\\
Ti&--0.23&...&1&--0.31&--0.52&--0.31&--0.25&0.28&--0.12\\
S&--0.43&...&1&...&--0.41&...&...&0.08&--0.35\\
Al&0.00&...&1&...&--0.32&--0.21&--0.18&0.51&--0.18\\
Ni&--0.46&0.13&4&--0.49&--0.51&--0.46&--0.46&0.04&--0.49\\
Na&--0.38&0.10&1&...&--0.61&...&--0.41&0.13&--0.52\\
Mn&--0.51&0.01&2&--0.89&--0.59&--0.53&--0.73&0.00&--0.53\\
K&--0.34&...&1&...&--0.54&--0.29&--0.32&0.17&--0.32\\
Cr&--0.67&...&1&--0.58&--0.58&...&--0.57&--0.53&--0.58\\
Co&--0.78&...&1&--0.41&...&--0.42&--0.43&--0.27&--0.41\\
V&--0.45&...&1&--0.44&--0.77&--0.39&--0.32&0.06&--0.32\\
Cu&--0.89&...&1&...&...&--0.66&...&--0.37&--0.66\\

\hline
 & & & &  & Non-Differential & & & \\
 \hline
Fe&--0.54&0.08&20&--0.52&--0.58&--0.47&--0.52&0.00&...\\
P&--0.22&...&1&...&...&...&...&0.29&...\\
Rb&...&...&1&...&...&...&...&...&...\\
Yb&...&...&1&...&...&...&...&...&...\\

\hline \hline
\end{tabular}
\\ \\
\tablefoot{The chemical abundances derived for Arcturus from the BACCHUS pipeline with N lines are shown in Column 2 and 3, respectively. The literature abundances are taken from ($^{a}$) \cite{Jofre2015}, ($^{b}$) \cite{Holtzman2015}, ($^{c}$) \cite{Smith2013} and ($^{d}$) \cite{Ramirez2011} and can be found in columns 4, 5, 6, and 7, respectively. The adopted [X/H] value for Arcturus for the elements used in the differential analysis are given in Column 10. }
\label{tab:Arcturuschem}
\end{table*}

\section{Results}
\label{sec:results}
In this section, we present the results of the analysis starting with the metallicity and \vmic\ parameters (Sect. \ref{subsec:SP}). We then present the results of the chemical abundance analysis for the APOKASC sample and turn to a discussion, on an element-by-element basis, of the quality of the measurements in Sect. \ref{subsec:Chemistry}. Finally, in Sect. \ref{subsec:clusterchem} we describe the procedures that were used to estimate the precision in the chemical abundance ratios. 

\subsection{Metallicity}
\label{subsec:SP}
\begin{figure}
	 \includegraphics[width=0.95\columnwidth]{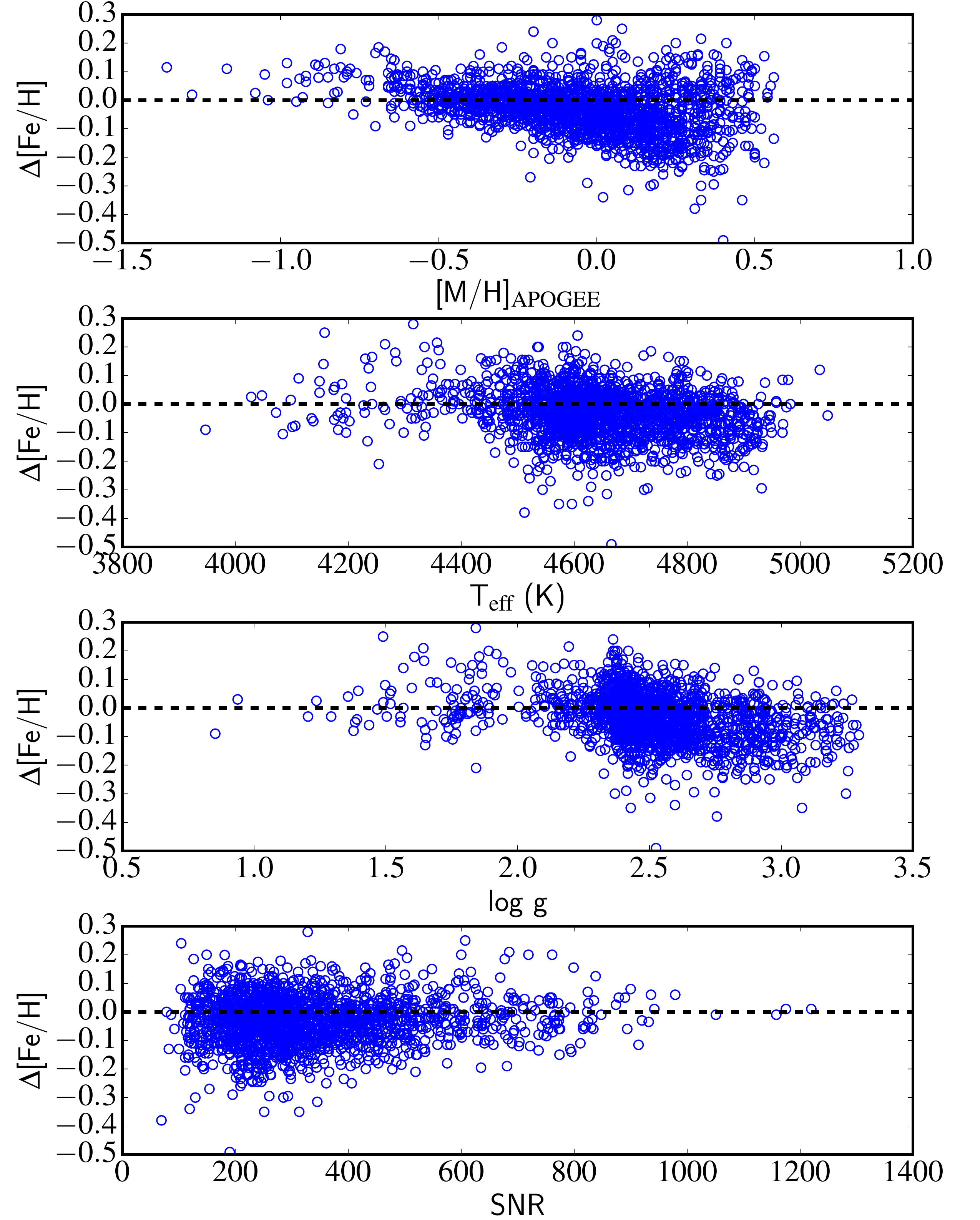}
	\caption{The difference in the \feh\ derived from BACCHUS and the corrected [M/H] from APOGEE, $\Delta$\feh\ = \feh$_{\mathrm{BACCHUS}} -$ [M/H]$_{\mathrm{APOGEE}}$, as a function of the APOGEE DR12 corrected global metallicity (denoted as [M/H]$_{\mathrm{APOGEE}}$), \teff, \logg, and SNR for the APOKASC sample from top to bottom, respectively. }
	\label{fig:fehdel}
\end{figure}

Fig. \ref{fig:fehdel} shows the difference in metallicity determined from BACCHUS and the global corrected [M/H] value from APOGEE as a function of APOGEE DR12 corrected [M/H], \teff, \logg, and SNR from top to bottom, respectively. There is good agreement between the results from BACCHUS and that of the corrected [M/H] from ASPCAP for these stars. The typical difference between our \feh\ and the calibrated [M/H] from APOGEE is --0.03 dex with a dispersion of 0.08 dex (which is of the order of the line-to-line abundance dispersion for Fe). There does not seem to be a strong trend in the difference of the corrected [M/H] and the BACCHUS derived [Fe/H] as a function of \teff\ and SNR. However, there is a weak but significant trend with metallicity and \logg\ such that the BACCHUS \feh\ are slightly lower than the corrected [M/H] at the high metallicity (and \logg) end. As suggested in Fig. \ref{fig:clusters}, this could be a result of the calibration applied to the ASPCAP pipeline results.

\subsection{Microturbulence} \label{subsec:vmic}
\cite{Smith2013} made use of 13 \FeI\ lines with a range of similar line strengths as this study. However, in this study we increase the number of lines from 13 to 20 in order to better quantify the correlation between abundance and REW to constrain \vmic. This contrasts the procedures followed by APOGEE DR12. 
\begin{figure}
	 \includegraphics[width=1\columnwidth]{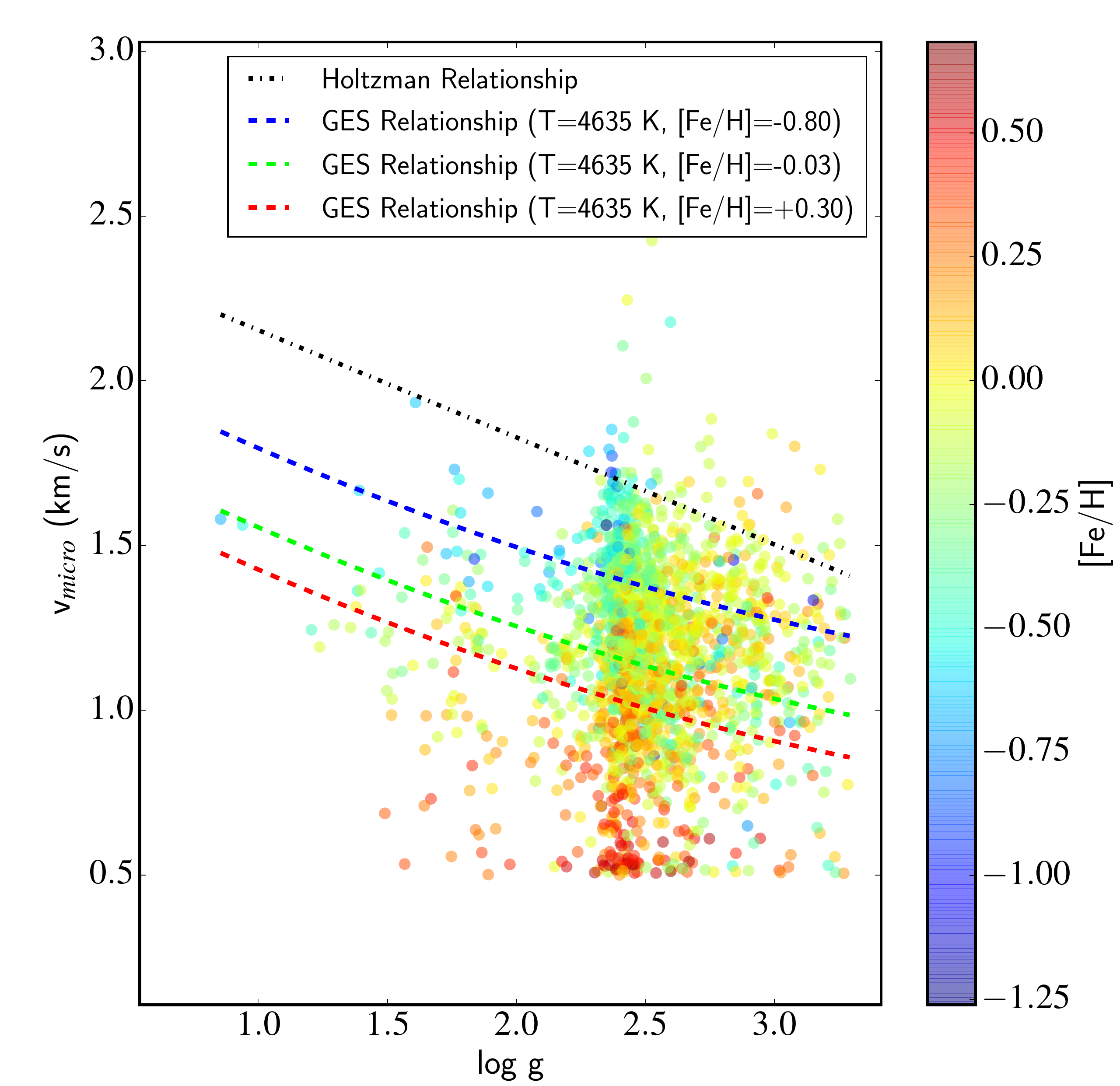}
	\caption{The derived \vmic\ as a function of \logg\ for the APOKASC sample. The color-code represents \feh. The black dot dashed line represents the \vmic-\logg\ relationship given in Equation 1 of \cite{Holtzman2015}. The blue, lime and red dashed lines represent the \vmic-\logg\ relationship from the GES at a \teff = 4635~K (median \teff\ of the sample), and \feh = --0.80, --0.03, +0.30 dex, respectively.  }
	\label{fig:vmiclogg}
\end{figure}

\cite{Holtzman2015} derived \vmic\ for a subsample of the data and used that as a basis to construct an empirical relationship between \logg\ and \vmic\ that they apply to the full sample (see their Sect. 4.2). However, Fig. 2 of \cite{Holtzman2015} suggests that there is more than just a \logg\ dependence for \vmic. In addition to the line selection (see Sect. \ref{subsec:lineselection}), it is possible that one reason APOGEE DR12 overestimates \feh\ of stars at low metallicity is due to the discrepancy between the derived and assumed \vmic\ in this regime \citep[Fig. 2 of ][]{Holtzman2015}. Indeed, this is why GES uses an empirical relation for \vmic\ that relates, in addition to \logg, the \teff\ and metallicity of the star \citep[e.g.][]{Smiljanic2014}. In Fig. \ref{fig:vmiclogg}, we show the derived \vmic\ of the APOKASC sample as a function of \logg\ color-coded by \feh. We also show the empirical \vmic-\logg\ relationships from \cite{Holtzman2015} and the GES \citep[][]{Smiljanic2014}. The \vmic\ determined from BACCHUS is offset with those adopted by APSCAP. It is also interesting to note that the GES relationship indicates that at the median \teff\ of the sample (\teff = 4635~K), the overall \vmic\ decreases with increasing metallicity which is exactly what is observed with the results from BACCHUS. The lowest metallicity stars have, on average, higher \vmic. We note here that for a small number of stars ($\sim$4\% of the sample) we flag as suspicious because they either have \vmic\ larger than 2.5 \kms\ or less than 0.60 \kms.

\begin{figure}
	 \includegraphics[width=0.97\columnwidth]{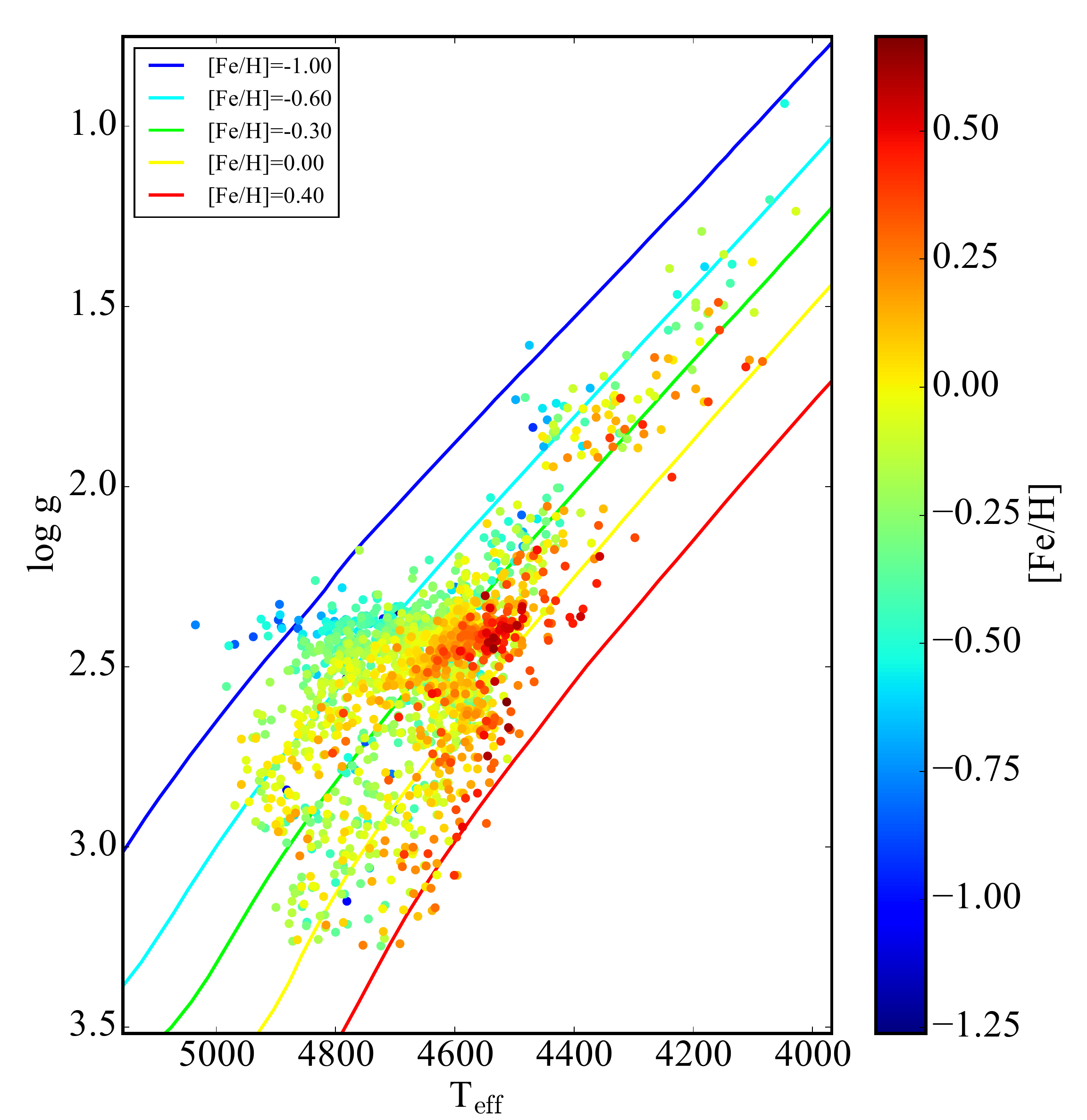}
	\caption{\logg\ as a function of \teff\ for the APOKASC sample. Overlaid are five 5-Gyr Y$^2$ isochrones spanning in metallicity from --1.0 $<$ \feh\ $<$ +0.4 dex. }
	\label{fig:HRD}
\end{figure}

Another common way to assess parameters is to recover the expected trends in the HR diagram. Therefore, in Fig. \ref{fig:HRD} we plot \teff\ as a function of \logg\ color-coded by metallicity. We have over laid tracks from the Yonsei-Yale\footnote{http://www.astro.yale.edu/demarque/yystar.html} isochrones \citep[Y$^2$,][]{Yi2003,Demarque2004}. Fig. \ref{fig:HRD} indicates that the stars follow the expected trend in \teff-\logg-\feh\ space and thus we conclude that no post-calibrations on the stellar parameters are needed.

\subsection{APOKASC Chemical Abundances}
\label{subsec:Chemistry}

\begin{figure*}
	 \includegraphics[width=2\columnwidth]{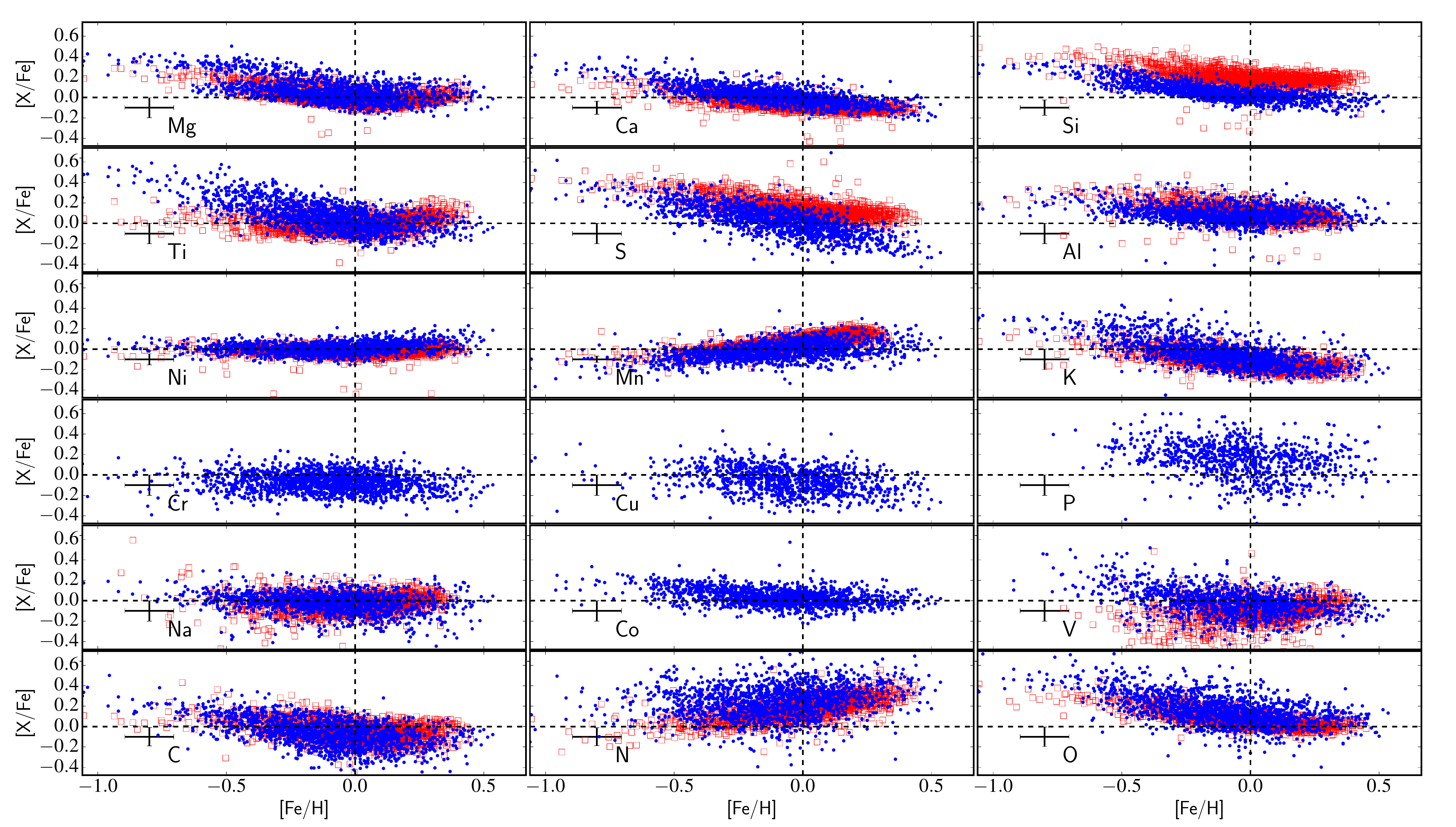}
	\caption{The [X/Fe]-[Fe/H] diagram for each element for the APOKASC sample from the BACCHUS pipeline (blue points). We also plot the [X/Fe] derived from the ASPCAP (red open squares) for 18 elements. The error bars represents the median uncertainty in [Fe/H] and [X/Fe].}
	\label{fig:allchem_APOGEE}
\end{figure*}
One of the primary advantages of large spectroscopic surveys, such as APOGEE, is that homogeneously derived chemical abundances can be determined for a large number of stars which enables bulk chemical abundance studies of the Milky Way. Thus one of the goals of this work is to provide updated chemical abundances, with minimal calibrations, for the nearly 2000 stars in the APOKASC catalogue. Now that the stellar parameters have been determined, we derived the abundances of up to 21 additional elements. These elements include Mg, Ca, Si, Ti, S, Al, Ni, Na, Mn, K, Cr, Co, Cu, P, Rb, Yb, and V. However, they do not include Fe, C, N, O which are determined along with the stellar parameters (see Sect. \ref{subsec:BACCHUS} for more details). 

In Fig. \ref{fig:allchem_APOGEE}, we present the [X/Fe]-[Fe/H] diagram for each element (including CNO) for the APOKASC sample derived from BACCHUS with the ASPCAP values in the background. We further remind the reader that all elements except P and Fe are derived using a line-by-line differential analysis and are solar scaled using the [X/H] values of Arcturus in Table \ref{tab:Arcturuschem}. The other elements listed are solar scaled by using the solar abundances of \cite{Asplund2005}. We do not display the results for Rb, Yb, and Cu because, as it will be shown below, they only represent upper limits at this time.

The top five of panels from Fig. \ref{fig:allchem_APOGEE} indicate that we recover the expected trends in the $\alpha$-elements with respect to \feh, namely that at low metallicity (\feh\ $<$ --0.80 dex) there is a plateau at high [Mg, Ti, Si, Ca, S/Fe] and at higher metallicities there is a decrease in those abundance ratios towards increasing metallicity. Furthermore, those panels indicate that we reproduce the precision of the ASPCAP pipeline. However, there are minor differences in the range of certain abundance ratios comparing BACCHUS to ASPCAP. For example, while [Mg/Fe] shows very good agreement between the two pipelines, at lower metallicities the BACCHUS [Mg/Fe] plateaus at a higher value, near +0.40 dex, compared to ASPCAP. 

We also found that the [Si/Fe] ratios derived with BACCHUS are on the order of 0.20 dex lower than the ASPCAP values. It is noted in the literature \citep{Holtzman2015, Masseron2015, Hawkins2015b}, that there are possible issues with the accuracy (i.e. zero-point) of ASPCAP abundances (particularly S, Si, and N). Additionally, the [Ti/Fe] panel indicates that the derived [Ti/Fe]-metallicity relationship from this study is similar to optical studies of nearby stars \citep[e.g.][]{Bensby2014}. On the other hand, the ASPCAP abundance results show increasing [Ti/Fe] with increasing [Fe/H] inconsistent with the literature. This is, in large part, why \cite{Holtzman2015} flags Ti as unreliable. We have solved this with our line selection. It is clear in Fig. \ref{fig:tilines} that the [Ti/Fe]-[Fe/H] pattern seen in APOGEE is found in the deselected lines at 15715.6~\AA\ and 15334.8~\AA.

The abundance ratios of [Ni/Fe], [Al/Fe], and [Mn/Fe] are in good agreement with the APOGEE DR12 results. However, there are some differences between DR12 and BACCHUS. In particular, the [Al/Fe] ratios derived in this study are smaller at low metallicity compared to DR12 at the 0.07 dex level. [Mn/Fe] abundance ratios from BACCHUS are lower by as much as $\sim$0.10 dex at high metallicity compared with APOGEE DR12 but the two are in excellent agreement at low metallicities. 

Both the [K/Fe] and [Na/Fe] are in fair agreement with APOGEE DR12. At low metallicity, the [K/Fe] derived using the BACCHUS pipeline is larger by $\sim$0.10~dex compared to the values from APOGEE DR12. However, these values tend to agree at metallicities larger than --0.30 dex. The [V/Fe] shows an increasing trend with decreasing metallicity, in contrast to what is found in \cite{Holtzman2015}. [V/Fe] is currently not recommended by APOGEE DR12 because it displays a large scatter. However, as we will show in Sect. \ref{sec:discussion}, we found that the [V/Fe]-[Fe/H] pattern found in this study is consistent with the literature. The [C/Fe] is in good agreement with the ASPCAP values but [N/Fe] derived in this study are $\sim$0.10 dex higher compared with ASPCAP. Similarly, the [O/Fe] is offset compared to APOGEE DR12 at low metallicity by +0.20~dex but becomes in good agreement at high metallicity. 
\begin{figure}
	 \includegraphics[width=\columnwidth]{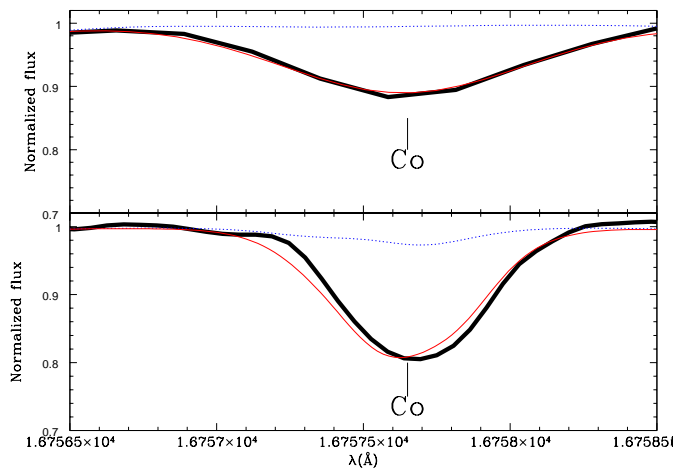}
	\caption{Detection of a Co line in Arcturus. The thick black line is the observation, the thin red line is the synthesis with Co and the dotted blue line is the synthesis without the Co line. The upper panel shows the line at the APOGEE resolution while bottom panel shows the line at high-resolution.}
	\label{fig:Co_line}
\end{figure}
\begin{figure}
	 \includegraphics[width=\columnwidth]{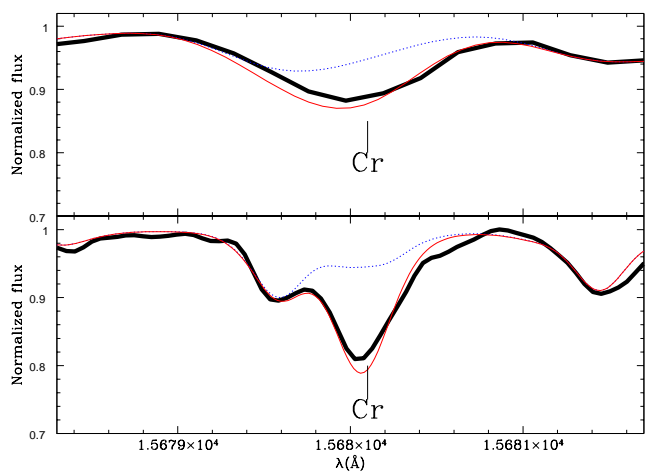}
	\caption{As in Fig \ref{fig:Co_line}, this time showing for Cr detection. }
	\label{fig:Cr_line}
\end{figure}

Co and Cr are new elements not in the DR12 release, and we present in Fig.~\ref{fig:Co_line} and  Fig.~\ref{fig:Cr_line}, the lines that have been used as they appear in Arcturus. Fig.~\ref{fig:Co_line} indicates that they are both strong and only weakly blended. There is good agreements between the model and observed spectra at both high- and moderate-resolution. Fig.~\ref{fig:Cr_line} indicates that the Cr line is slightly blended, with CN, at bluer wavelengths. However, the synthesis fit is adequate at both resolution settings because of our good CN line list. We also note that the Cr line is in a region heavily blended by telluric features but these seem to be well subtracted in the ASPCAP reduction pipelines.

\begin{figure}
	 \includegraphics[width=\columnwidth]{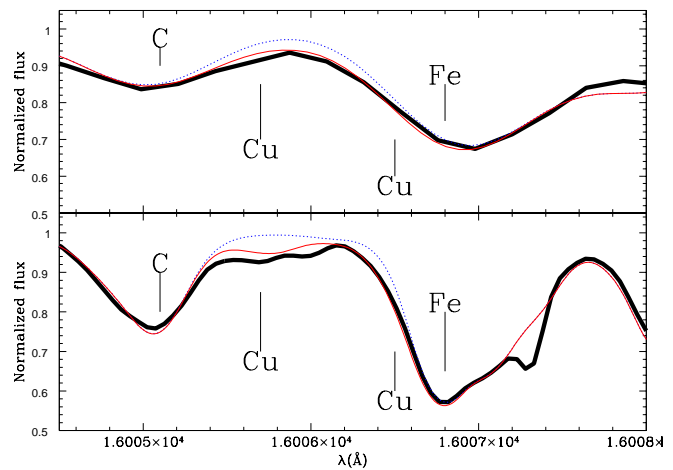}
	 \caption{Detection of two Cu lines. Line styles as in Fig \ref{fig:Co_line}. }
	\label{fig:Cu_line}
\end{figure}

\begin{figure}
	 \includegraphics[width=\columnwidth]{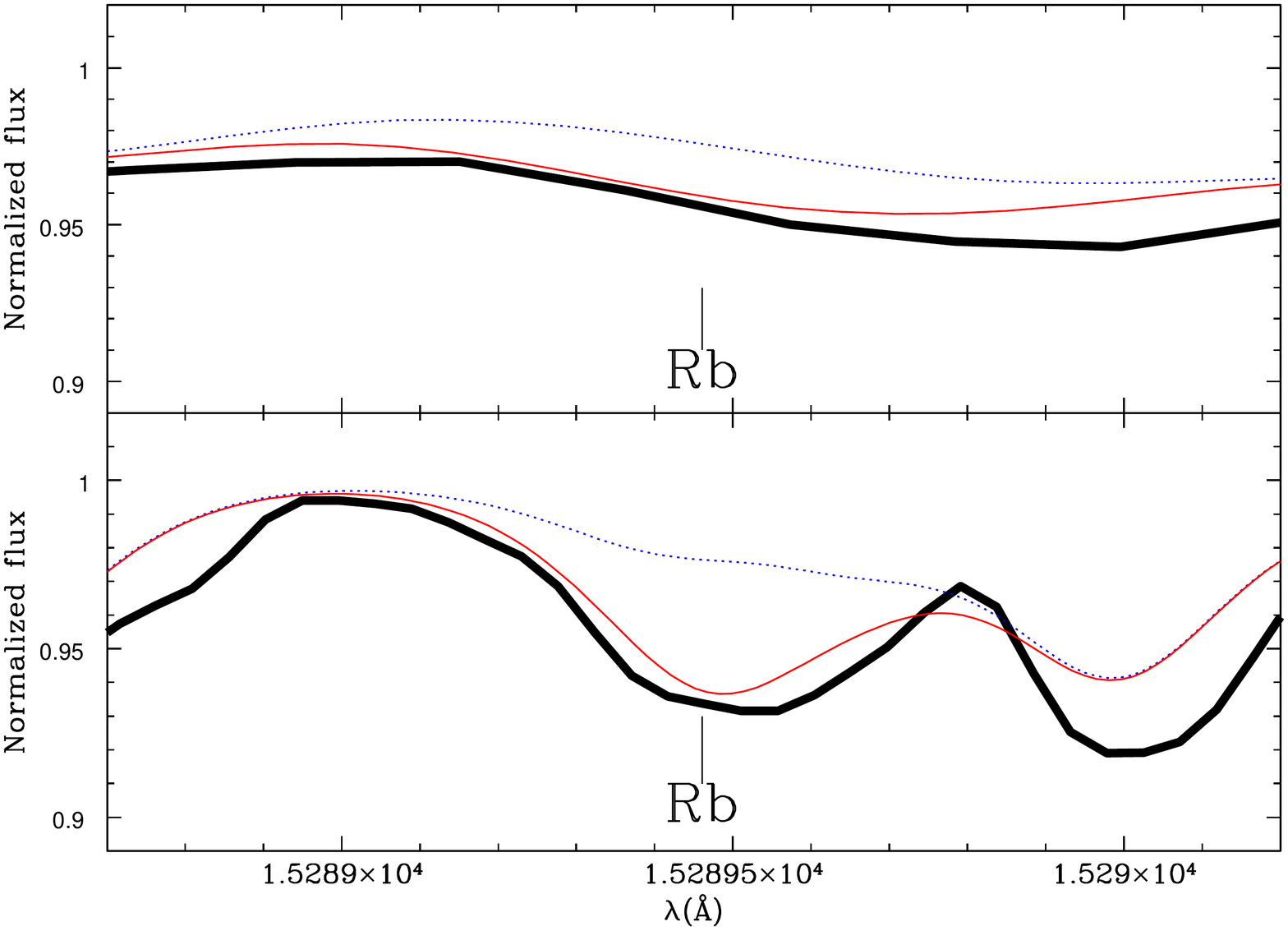}
	 \caption{Detection of a Rb line. Line styles as in Fig \ref{fig:Co_line}. }
	\label{fig:Rb_line}
\end{figure}

\begin{figure}
	 \includegraphics[width=\columnwidth]{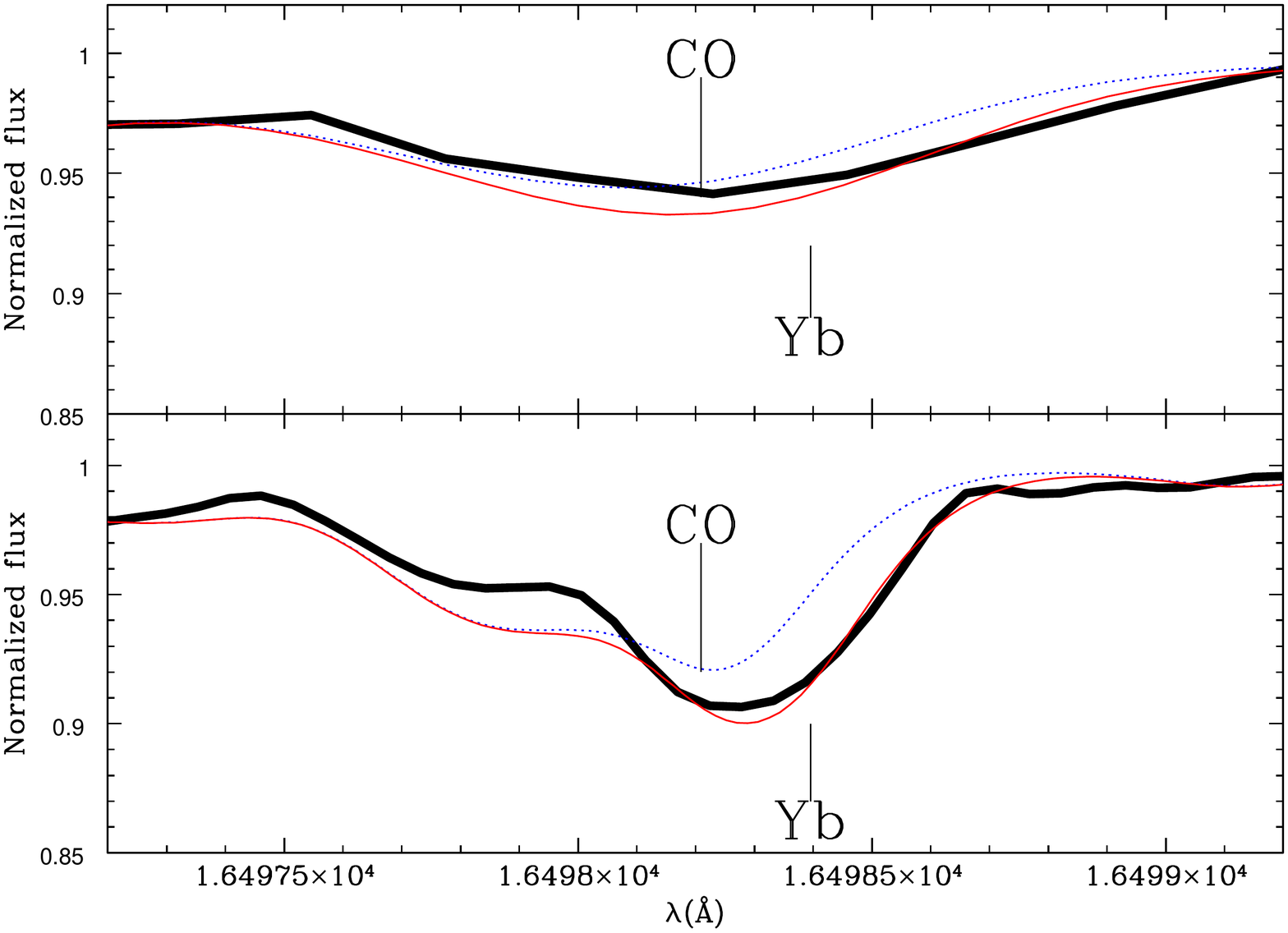}
	 \caption{Detection of a Yb line. Line styles as in Fig \ref{fig:Co_line}. }
	\label{fig:Yb_line}
\end{figure}

\begin{figure}
	 \includegraphics[width=\columnwidth]{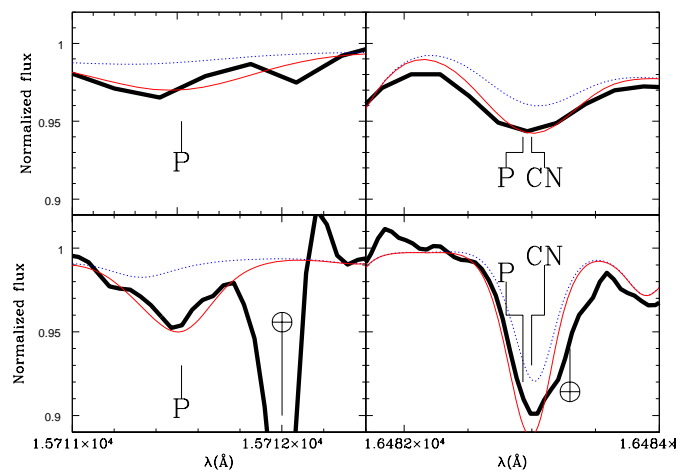}
	 \caption{Detection of two P lines. Line styles as in Fig \ref{fig:Co_line}.  In both panels the $\bigoplus$ symbol represents a blend from a telluric feature.}
	\label{fig:P_lines}
\end{figure}

We have also attempted to measure an additional four new elements, namely Cu, Rb, Yb, and P, because they have detected lines in the Arcturus spectrum (Fig.~\ref{fig:Cu_line}, \ref{fig:Rb_line}, \ref{fig:Yb_line}, and \ref{fig:P_lines}). However, those lines are weak and heavily blended and therefore the derived abundances strongly depends on the ability of properly reproducing the blend. At this time, we cannot guarantee the quality of the abundances for those elements. However, while Rb and Yb are extremely challenging, Cu and P both present two promising and strong lines. In Sect. \ref{sec:discussion}, we compare these results with literature.



\subsection{Chemical Abundance Precision} \label{subsec:clusterchem}
Evaluating the precision of the chemical abundance can be done in several ways. One way is by deriving the star-to-star dispersion of the chemical abundances within a set of validation globular and open clusters. This works because clusters are thought to be chemically homogenous in many elements \citep[e.g.][]{Holtzman2015, Bovy2016}. However, it is important to note that with anti-correlations, such as the Na-O or Mg-Al, specific elements (e.g. C, N, O, Mg, Al, Na) have been shown to be variant within clusters \citep[e.g][and references therein]{Gratton2001, Gratton2012}. However, we proceed under the assumption, for this test only, that globular and open cluster are chemically homogenous in all elements except CNO.

\begin{figure*}
\centering
	 \includegraphics[width=2\columnwidth]{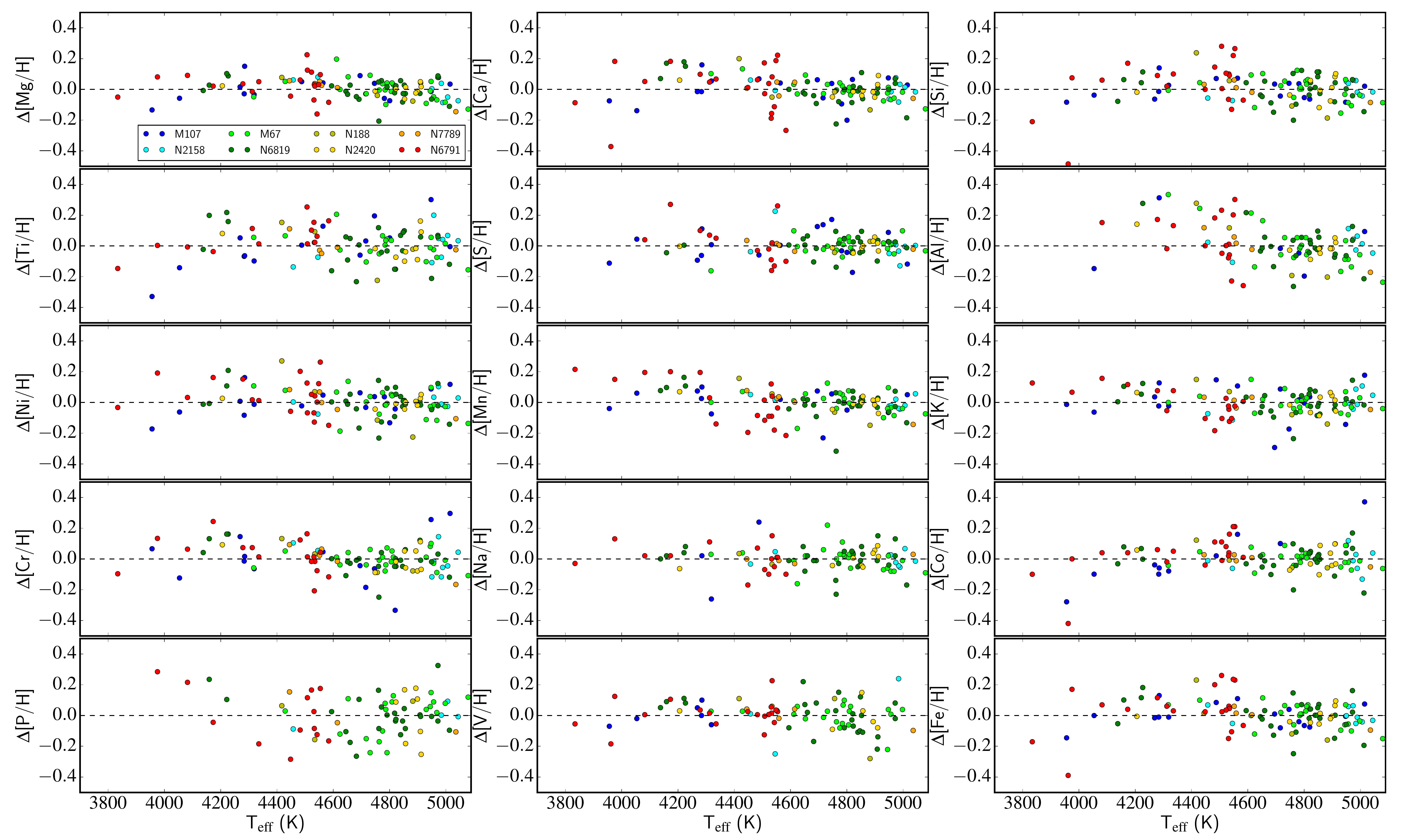}
	\caption{The difference between the [X/H] for each star in a cluster and the cluster mean, $\Delta$[X/H] = [X/H]$_{\mathrm{star}} -  <$ [X/H]$_{\mathrm{cluster}}>$, as a function of \teff\ for every stars within each validation cluster. The color-coding indicates the cluster that the star belongs to (e.g. all blue points are from the metal-poor cluster M107). } 
	\label{fig:clusterabu}
\end{figure*}

If this is the case, the precision with which one can measure chemical abundances can be assessed by deriving the star-to-star dispersion of each element within a cluster. In addition, the elemental abundances should not depend on a star's stellar parameters within the cluster. We have measured the abundances described in Sect \ref{subsec:Chemistry}, in all globular and open cluster used for validation in the same way as the APOKASC sample. We remind the reader that unlike the APOKASC stars, most of these stars do not have astroseismic \logg\ information and thus could have larger uncertainties in the \logg\ and hence larger abundance uncertainties. In Fig. \ref{fig:clusterabu} we show the difference of each star's [X/H] abundance ratio subtracted from the cluster's mean [X/H] ratio as a function of \teff. If the clusters were chemically homogenous and we could measure the abundances with infinite precision then each point in Fig. \ref{fig:clusterabu} would have $\Delta$[X/H] = 0. Thus the dispersion around zero yields an approximation of the total internal uncertainty in the abundances. In column 2 of Table \ref{tab:abunderror}, we show the typical star-to-star dispersion of [X/H] (around the mean value) within the validation clusters. These values are typically around $\sim$0.10 dex and are similar to the uncertainties reported by APOGEE.

\begin{table}
\setlength{\tabcolsep}{3pt}
\caption{Typical Abundance Uncertainties in Clusters and Sensitivity.} 
\begin{tabular}{c c c c c c} 
\hline\hline
[X/H]&$\sigma$[X/H]$_{\mathrm{clus}}$& $\Delta$\teff &$\Delta$\logg&$\Delta$\vmic & $\sigma_{\mathrm{line}}$\\
 & & ($\pm$80~K)& ($\pm$0.02)&($\pm$0.1~\kms) &\\
\hline\hline
Mg & 0.07& $\pm$0.05&$\pm$0.01&$\mp$0.01&0.04\\
Ca & 0.09&$\pm$0.06&$\pm$0.03&$\mp$0.01&0.03\\
Si & 0.10&$\pm$0.03&$\pm$0.01&$\mp$0.03&0.04\\
Ti&0.11&$\mp$0.07&$\pm$0.01&$\mp$0.02&...\\
S&0.07&$\mp$0.04&$\pm$0.01&$\mp$0.01&...\\
Al&0.15&$\pm$0.10&$\pm$0.02&$\pm$0.01&...\\
Ni&0.11&$\pm$0.05&$\pm$0.00&$\mp$0.01&0.03\\
Mn&0.09&$\pm$0.08&$\pm$0.00&$\mp$0.02&...\\ 
K&0.08&$\pm$0.10&$\pm$0.01&$\pm$0.01&...\\
Cr&0.10 & $\pm$0.11 & $\pm$0.01&$\mp$0.02&...\\
Na&0.09&$\pm$0.07&$\pm$0.02&$\mp$0.01&...\\
Co&0.09&$\pm$0.05&$\pm$0.00&$\mp$0.01&...\\
V&0.10&$\pm$0.10&$\pm$0.00&$\mp$0.01&...\\
Fe&0.08&$\pm$0.07&$\pm$0.02&$\mp$0.04&0.02\\
C&...&$\pm$0.03&$\pm$0.03&$\mp$0.03&0.04\\
N&...&$\pm$0.06&$\pm$0.01&$\mp$0.01&0.02\\
O&...&$\pm$0.10&$\pm$0.03&$\mp$0.04&0.04\\


 \hline
 \end{tabular}
\\ \\
\tablefoot{The typical star-to-star dispersion in [X/H] within the eight validation clusters is displayed for every element in column 2. Typical sensitivity of the abundance to uncertainties in the stellar parameters are computed by measuring the difference in [X/H] abundance due to changes of $\pm$80~K (column 3), $\pm$0.02 dex (column 4), and 0.10~\kms\ (column 5), in \teff, \logg\ and \vmic, respectively. This was done  using three stars near the middle of the parameter space (i.e. \teff\ $\sim$ 4700~K, \logg\ $\sim$ 2.5, \feh\ $\sim$ --0.2). The standard error, $\sigma_{\mathrm{line}}$, of the mean abundance is displayed in column 6 for elements where there are more than two lines. }
\label{tab:abunderror}
\end{table}

We note that as in \cite{Holtzman2015}, we could have chosen to internally calibrate the [X/H] abundances by fitting out any correlations between abundance and \teff. We have chosen not to do this because (1) we have only 119 stars which are dominated by 4600 $<$ \teff\ $<$ 4800~K, thus we may not adequately sample the low temperature regime, (2) we do not have a good handle on the intrinsic abundance spreads within clusters particularly with the known anti-correlations, (3) the slope of the regression fit of $\Delta$[X/H] and \teff\ tends to vary significantly by removing just one cluster indicating that the calibration would not likely be robust.

Another way of evaluating the total internal uncertainty in the chemical abundances is to quantify the sensitivity in the abundance due to the uncertainties in the stellar parameters (\teff, \logg, and \vmic) and the mean abundance. We selected three stars for this analysis: (1) 2MASS J19070835+5016440 with \teff\ = 4882~K, \logg\ = 2.99~dex, \feh\ = --0.29~dex, (2) 2MASS J18583782+4822494 with \teff\ = 4740~K, \logg\ = 2.54~dex, \feh\ = --0.08~dex, and (3) 2MASS J19103742+4934534 with \teff\ = 4689~K, \logg\ = 2.39~dex, \feh\ = --0.08~dex. These stars are a good representation of the APOKASC sample and thus it is reasonable to assume that their performance reflects the typical star analyzed in this study. For each of these stars we changed \teff\ by $\pm$80~K, \logg\ by $\pm$0.02~dex, which are the their typical uncertainties \citep{Pinsonneault2014}, and \vmic\ conservatively by $\pm$0.10 \kms. The abundance deviations due to the changes in the stellar parameters can be found in Table \ref{tab:abunderror}. The total internal uncertainty is then the abundance sensitivity to the stellar parameters added in quadrature with the standard error in the mean [X/H] abundance. We note that the uncertainty measured from the star-to-star dispersion of each element within the validation clusters, and the sensitivity of the stellar parameters are comparable. 

\section{Discussion}
\label{sec:discussion}
\begin{figure*}
	 \includegraphics[width=2\columnwidth]{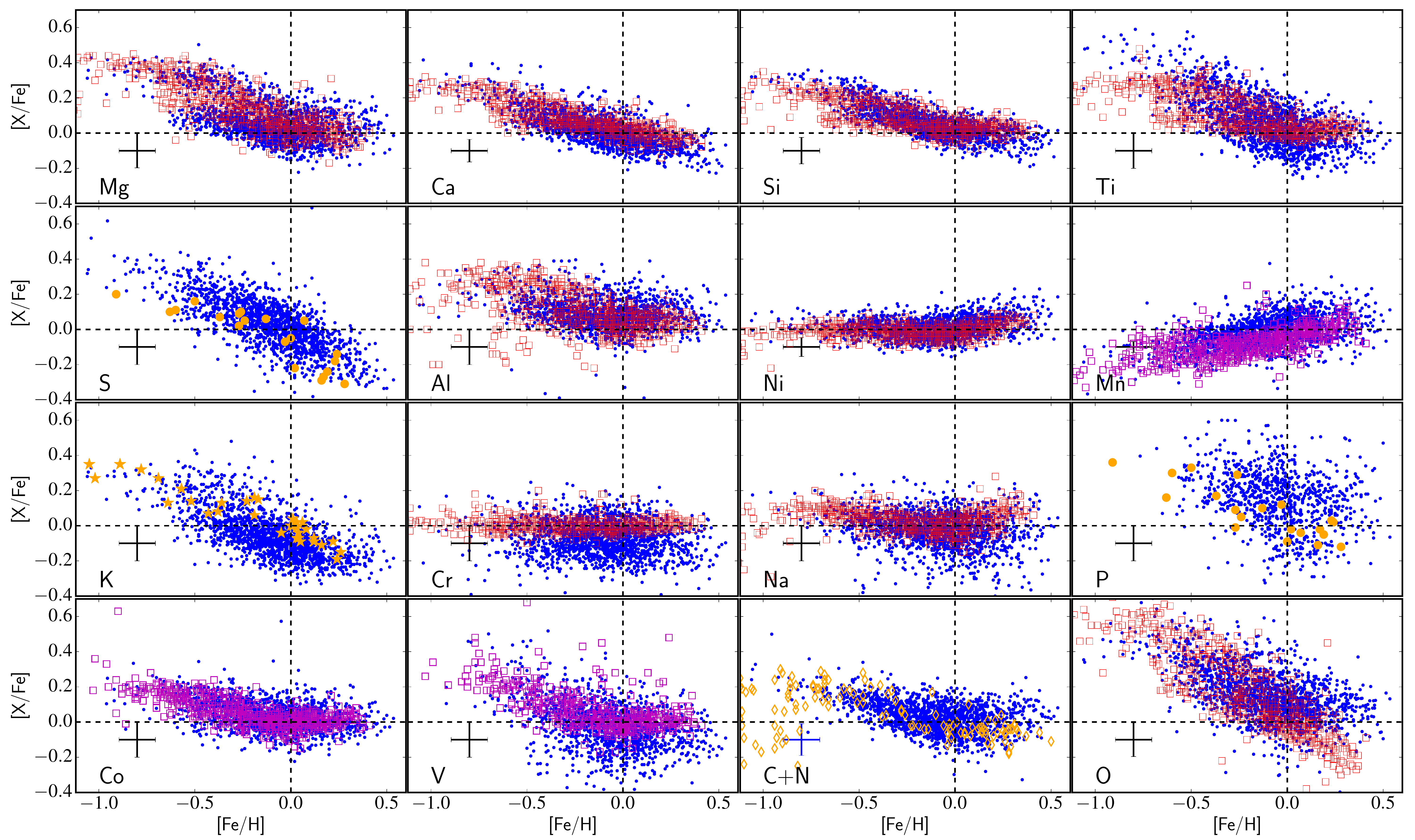}
	\caption{The [X/Fe]-[Fe/H] diagram for each element for the APOKASC sample from the BACCHUS (blue points) and a local sample of dwarf stars  observed in the optical from \cite{Bensby2014} denoted by red open squares and \cite{Battistini2015} denoted as open magenta squares. In addition, the [P/Fe]  and [S/Fe] are taken from \cite{Caffau2011} and are denoted as orange circle. The literature [K/Fe] are sourced from \cite{Shimansky2003} and are denoted as orange stars. The literature data for the [C+N/Fe] panel is the [C/Fe] data taken from \cite{Nissen2014} and is denoted as orange open diamonds. The error bars represents the median uncertainty in [Fe/H] and [X/Fe].}
	\label{fig:abundlit}
\end{figure*}

In this section, we discuss, element-by-element, how the results compare with a local sample of stars often observed using high-resolution optical spectra. For this, we refer the reader to Fig. \ref{fig:abundlit} which presents the [X/Fe] values derived in this study as a function of metallicity compared with several literature sources in the background. The literature data is taken from local samples of stars observed with high-resolution optical spectra. For most elements we draw from the work of \cite{Bensby2014} and \cite{Battistini2015}. However for [P/Fe] and [S/Fe] we take data from \cite{Caffau2011}. Additionally, [K/Fe] and [C/Fe] data are sourced from \cite{Shimansky2003} and \cite{Nissen2014}, respectively.

\subsection{The $\alpha$-elements: O, Mg, Si, S, Ca, and Ti} \label{subsec:alphael}
The $\alpha$-elements are critical to understanding the star formation history of the Galaxy. This is because the ratio of $\alpha$-elements to iron peak elements is linked to environment with which the population of stars of interest were born \citep[e.g.][]{Gilmore1998, Matteucci2001}. The mean ratio of these elements, \afe, has been show to be zero near solar metallicities. The \afe\ ratio tends to increase toward decreasing metallicity until \feh\ $\sim$--1.0 dex where it stabilizes around a plateau \citep[e.g.][]{Edvardsson1993,Reddy2003,Venn2004,Fuhrmann2004,Bensby2014}. Structurally, the thin disk of the Galaxy is thought to be comprised of metal-rich stars with low \afe, while the thick disk is thought to be more metal-poor with higher \afe\ compared to the thin disk. 

In the top five panels of Fig. \ref{fig:abundlit}, we observe the expected trend of increasing [Mg, Ca, Si, Ti, S, O/Fe] with decreasing metallicity. The abundance ratio of these elements are consistent with the literature. This is not the case for Ti, Si and S for ASPCAP. More specifically, at low metallicities, the plateau of [Mg, Ca, Si/Fe] derived in this work occurs at similar values as the literature. [S/Fe] is known, from the work of \cite{Caffau2011}, to decrease to subsolar values at high metallicities (\feh $>$ 0.00 dex) which is recovered here but not in APOGEE. In addition, we improve the zero-point issue seen in APOGEE (e.g. [Si/Fe]) by the implementing the differential analysis. 

Furthermore, we recover the expected [Ti/Fe]-[Fe/H] trend, albeit, with somewhat poorer precision compared to the high-resolution optical study of \cite{Bensby2014}. As mentioned before, this improvement is a result of the careful line selection. The lines which show [Ti/Fe]-\feh\ trends similar to APOGEE DR12 are those which are saturated or possibly strongly affected by NLTE (see Sect. \ref{subsec:lineselection} for more details). 

Interestingly, the [O/Fe] abundance ratio at super-solar metallicity has a slight discrepancy with \cite{Bensby2014}. The [O/Fe] derived both in this study, and in APOGEE DR12 shows a flat trend at super-solar metallicities. However, the [O/Fe] in \cite{Bensby2014} shows a decreasing trend with increasing metallicity at \feh\ $>$ 0.00~dex. This could be a result of NLTE effects. As noted in \cite{Bensby2004}, the downward trend in [O/Fe] with increasing metallicity at \feh\ $>$ 0.00 dex is seen for the forbidden [O I] line at 6363 \AA, which is known not to be strongly affected by NLTE effects. However, if the [O/Fe] is computed from the oxygen triplet at $\sim$ 7774~\AA, which is known to be strongly affected by NLTE, the trend at  \feh\ $>$~0.00 dex is significantly flatter, consistent with what we observe in Fig~\ref{fig:abundlit}. This issue shows the power of doing Galactic archaeology combining different data set, as this allows us to take the best of each of them.


\subsection{The Fe-peak elements:  Mn, Ni, Co, Cr} \label{subsec:fepeak}
The Fe-peak elements (Mn, Ni, Co) are produced and dispersed, in large quantities, in Type Ia supernovae similar to Fe but in contrast to the $\alpha$-elements \citep[e.g.][]{Iwamoto1999}. However, many Fe-peak elements are also produced in Type II supernovae \citep[e.g.][]{Kobayashi2006}.Therefore, the evolution of these elements are thought to scale with Fe. 

It is expected that [Mn/Fe] has a decreasing trend with decreasing metallicity. This trend is possibly a result of either a delay in the enrichment from Type Ia supernovae \citep[e.g.][]{Kobayashi2009} or metallicity dependent Mn yields from Type II supernovae \citep[e.g.][]{Feltzing2007}. The abundance determination of Mn is, however, complex due to hyperfine structure splitting. We have taken this into account in the line list. In addition, the [Mn/H] ratio of Arcturus, our reference star, ranges from --0.89 to --0.4~dex in the literature \citep[e.g.][]{Thevenin1998,Ramirez2011, Luck2005, Smith2013, Jofre2015}. Despite these complications, the expected behavior of [Mn/Fe] as a function of metallicity observed in Fig. \ref{fig:abundlit} matches a sample of local dwarf stars. However, there seems to be a very small offset, which causes our [Mn/Fe] to be too high on the 0.05 dex level compared to \cite{Battistini2015}. This offset could easily be due to the choice of [Mn/H] adopted for Arcturus. 

Ni and Cr track iron very well and thus are thought to vary essentially in lockstep with Fe. This was observed to be the case with the local sample of dwarf stars from \cite{Bensby2014}. The derived [Ni, Cr/Fe] have a near flat correlation with metallicity as seen in other studies. However, we note that the [Cr/Fe] derived in this study is significantly more dispersed compared to the results of \cite{Bensby2014}. In addition, [Ni/Fe] shows a slight upward trend with increasing metallicity at \feh\ $>$ 0.00~dex. This is also seen in the results of \cite{Bensby2014}.

Co is thought to be produced in both Type I and Type II supernovae \citep[e.g.][]{Kobayashi2006}. Its correlation with metallicity is not exactly flat, which may be explained through supernovae yields of Co that are both mass and metallicity dependent. Additionally, Co is known to be affected by hyperfine structure effects which are accounted for through the line list in this work. The [Co/Fe] in this study is in very good agreement with the local sample of dwarf stars from \cite{Battistini2015}. In addition, this is an element ASPCAP currently does not provide. 

\subsection{The Odd-Z and Light Elements} \label{subsec:lightevenodd}
\subsubsection*{C and N}
Carbon and nitrogen have been shown to be very important elements in red-giants as they can be used to aid in age determination \citep[e.g.][]{Masseron2015, Martig2016}. Carbon is made through several means but its production is thought to be dominated by He burning on the main sequence, Type II supernovae, and asymptotic giant branch (AGB) stars. In addition, C and N are affected by the dredge-up process, therefore, they do not stay constant over the lifetime of a star, unlike most elements \citep[e.g.][]{Iben1965}. However, the combined ratio of [C+N/Fe] is thought to remain constant over the lifetime of the star \citep[e.g.][]{Masseron2015, Hawkins2015b}. Therefore, in Fig. \ref{fig:abundlit}, we use [C+N/Fe] as a proxy for [C/Fe]. We compared our results of the [C+N/Fe] with the [C/Fe] values from \cite{Nissen2014}. There is very good agreement between the literature and the values derived from BACCHUS. 

\subsubsection*{Odd-Z elements: Na, Al, K, V, P, Cu}
Na and Al are odd-Z elements, which are produced in a variety of ways. Both are produced in carbon burning but Na is also produced in the NeNa cycle during H-shell burning and Al is also produced in the MgAl cycle \citep[e.g.][]{Samland1998}. Models predict that at \feh\ larger than --1.0 dex, [Na/Fe] and [Al/Fe] decrease with increasing metallicity. This has been shown in observations \citep[e.g.][]{Fulbright2000, Reddy2003, Bensby2014}. Fig. \ref{fig:abundlit} indicates that [Na/Fe] decreases with increasing metallicity, albeit with relatively large scatter. This is likely due to somewhat poor fitting of Na lines in the spectral data. [Al/Fe] also shows a decreasing trend with increasing metallicity. Additionally, there is good agreement in the [Al/Fe] abundance ratio derived in this work and in APOGEE. 

K is thought to be an odd-Z element formed in explosive oxygen burning during Type II supernovae \citep[e.g.][]{Samland1998}. Although the yields are still rather uncertain, the models predict that [K/Fe] decreases with increasing metallicity \citep[e.g.][]{Samland1998, Nomoto2013}. Fig. \ref{fig:abundlit} indicates that the abundance ratio of [K/Fe] has a decreasing trend with increasing metallicity. This result is consistent with expectations from both models and observational data \citep[e.g][]{Shimansky2003}. 

V is an odd-Z element produced in a similar mechanism as K and also through neon burning \citep[e.g.][]{Samland1998}. However, its yields are still uncertain. Models predict [V/Fe] to decrease with increasing metallicity at \feh\ $>$ --1.0 dex with values below zero. In Fig. \ref{fig:abundlit}, we find that [V/Fe] shows the expected trends and is consistent with the literature \citep[e.g.][]{Battistini2015}, however its values are positive instead of negative as expected by the model. This suggest the need of improvement in supernovae yields.

P is an odd-Z element thought to be produced during carbon and/or neon burning which is then released in Type II supernovae \citep[e.g.][]{Woosley1995, Samland1998, Caffau2013}. Until recently, there have been very few studies of P because there are no P lines in the typical wavelength range in observed spectra of FGK-type stars in the optical. While theoretical models predict [P/Fe] to be negative and decreasing toward increasing metallicity above \feh = --1.0~dex \citep[e.g.][]{Kobayashi2006}, the observations show mostly positive values for [P/Fe]. However, observations from \cite{Caffau2011} do confirm that there is a decreasing trend of [P/Fe] with increasing metallicity. We found that [P/Fe] globally decreases with increasing metallicity and is positive, consistent with the data from \cite{Caffau2011}. However, we remind the reader that the P line used in this study is in a region heavily affected by telluric features, as seen in Fig.~\ref{fig:P_lines}, which may be the reason for the large scatter in [P/Fe]. 

Cu is an odd-Z element, which is likely produced in a variety of ways. It is thought that the primary source is through Type Ia supernovae but it is also produced in secondary phenomena in massive stars, or weak s-processes \citep[e.g.][]{Mishenina2002}. Theoretically, it is expected that [Cu/Fe] shows a relatively flat trend with metallicity at \feh\ $>$ --1.0~dex and decreases with decreasing metallicity at \feh\ $<$ --1.0~dex \citep[e.g.][]{Kobayashi2006}. In Fig.~\ref{fig:allchem_APOGEE} [Cu/Fe] shows a roughly flat trend with metallicity with large scatter. As noted in Fig.~\ref{fig:Cu_line}, there are two heavily blended Cu lines which are detected. At least one line is blended with Fe and thus we cannot be sure that the flat trend with metallicity is astrophysical or due to the Fe blend. For this reason, and the significant scatter due to likely both telluric features and blending we caution on the accuracy of Cu but conclude that it is promising. 

\section{Summary}
\label{sec:conclusion}
In this paper, we have used an independent pipeline called BACCHUS described in Sect \ref{subsec:BACCHUS}, an updated line list and a careful line selection to explore the chemical abundance patterns of the APOKASC sample. In particular, we have been focused on solving the metallicity calibration issues pointed out, particularly at low metallicity in \cite{Holtzman2015} and adding additional elements to the abundance catalogue. We have selected the APOKASC sample as a first subsample to study because of the very precise \logg\ information that has been determined via astroseismology. As a result, we have fixed the \teff\ and \logg\ to those from \cite{Pinsonneault2014}, which has typical uncertainties on the order of 80~K, and 0.014 dex. We determined the remaining stellar parameters, metallicity and broadening parameters and chemical abundances of up to 21 elements. The results of this analysis can be summarized in the following points: 
\begin{enumerate}

\item In Sect \ref{subsec:Validation} we have shown that with BACCHUS and APOGEE spectra, we can accurately recover the metallicity of the three benchmark stars, the Sun, Arcturus, and \muleo, and 119 stars residing in eight globular and open clusters ranging in metallicity between --1.03 and +0.37 dex. This indicates that we do not need to calibrate the metallicity down to \feh\ = --1.0 dex. This is a significant improvement compared to the calibration required by APOGEE, which can be as large as 0.20 dex and can cause issues with [X/Fe] abundance ratios. We believe that this result was achieved through a careful line selection and solving for the broadening parameters (e.g. \vmic). We recommend that surveys fixing the \vmic\ using a relationship with \logg\ should also include \feh\ and \teff\ effects on that relationship.

\item We have provided a self-consistent and accurate set of abundances for up to 21 elements including C, N, O, Mg, Ca, Si, Ti, S, Al, Na, Ni, Mn, Fe, K, P, Cr, Co, Cu, Rb, Yb and V which, can be used as a training set for other spectral analysis procedures. The [X/Fe] abundance ratios and line-by-line abundances of these elements can be found in the provided online tables. Among these elements, there are two (Co and Cr), which are new compared to APOGEE DR12 and accurate in this study. There are also four additional new elements (Rb, Yb, Cu, and P) provided, which we caution as they are either heavily blended or display large scatter, which may be due to telluric features. 

\item We have shown the importance of line selection in chemical abundance analyses. In particular, through a careful line selection (i.e. deselecting saturated lines, or those, which are poorly reproduced in a high-resolution Solar and Arcturus spectrum) we have improved certain abundance ratios, e.g. [Ti/Fe], [V/Fe], which now follow the expected trends with metallicity found in the literature. We have presented a unique and powerful diagnostic diagram in Sect. \ref{subsec:lineselection} which has allowed us to discuss the impact of line selection on the final abundances. Using these diagrams, we have been able to diagnose at least one reason why APOGEE DR12 [Ti/Fe] abundance ratios show inconsistent trends with metallicity compared to the other studies. We have illustrated in Fig.~\ref{fig:tilines} that the selection of different lines (Ti in that case) by surveys can completely change the chemical patterns seen in the Galaxy. Therefore, we urge that surveys consider publishing line-by-line abundances which may allow the impact of line selection to be fully studied. 
\end{enumerate}

With these new, self-consistent, and accurate abundances it is possible to study chemical abundance trends in the outer Galaxy where the Kepler field resides, which is something we plan to discuss in a forthcoming paper.  We also point out that these new abundances are derived for stars with very precise \logg\ and \teff\ which improves their overall accuracy. Additionally, the line-by-line differential analysis has helped correct some zero-point offsets (e.g. in Si, S, and N) which make these abundances a superior set to train machine-learning style parameter and abundances solvers such as  {\it The Cannon} \citep{Ness2015, Casey2016}. Ultimately, the lessons learned in this study regarding solving for the broadening parameters, line selection, and line-by-line differential analysis can be incorporated in future APOGEE data releases and APOGEE2 when it comes online. 

\begin{appendix}
\section{Online Tables}
\label{app:A}

We provide our results in four online tables found in electronic format at the CDS. The first online table, a sample of which can be found in Table~\ref{tab:online_lines}, contains the absolute ($\log_\epsilon(X)$) abundances for every element and star on a line-by-line basis. We remind the reader that some of the $\log_\epsilon(X)$ values may be affected by improper line list data which is the motivation for the differential analysis. In addition, this table also contains the method-to-method scatter (see Sect. \ref{subsec:BACCHUS} for more details), which can be used to generate the diagnostic line-by-line abundance plot found in Fig. \ref{fig:tilines}. 
\begin{table}[ht]
\caption{Line-by-line Abundances Online Table Format.} 
  \setlength{\tabcolsep}{3pt}

\begin{tabular}{c c c c c}
\hline\hline
APOGEE ID & Element & $\lambda$ (\AA) & $\log_\epsilon(X)$ & e$\log_\epsilon(X)$ \\
\hline\hline
J18582020+4824064 & Al & 16763.4 & 6.460 & 0.084 \\
J18582020+4824064 & P & 15711.5 & 5.430 & 0.235 \\
J18582020+4824064 & S & 15478.5 & 6.649 & 0.093 \\
J18582020+4824064 & K & 15168.4 & 4.640 & 0.047 \\
J18582020+4824064 & Ca & 16150.8 & 5.988 & 0.076 \\
J18582020+4824064 & Ca & 16157.4 & 6.058 & 0.210 \\
J18582020+4824064 & Ti & 15873.8 & 4.568 & 0.196 \\
J18582020+4824064 & V & 15924.8 & 3.411 & 0.316 \\
J18582020+4824064 & Cr & 15680.1 & 5.159 & 0.085 \\
J18582020+4824064 & Mn & 15262.4 & 5.010 & 0.070 \\
J18582020+4824064 & Mn & 15159.2 & 4.953 & 0.137 \\
J18582020+4824064 & Ni & 15632.6 & 6.070 & 0.806 \\
J18582020+4824064 & Ni & 16584.5 & 6.170 & 0.045 \\
J18582020+4824064 & Ni & 16363.1 & 5.950 & 0.093 \\
J18582020+4824064 & Ni & 15555.4 & 5.880 & 0.152 \\
J18582020+4824064 & Cu & 16005.5 & 3.490 & 39.377 \\
J18571019+4848067 & Fe & 15194.5 & 7.400 & 0.195 \\
J18571019+4848067 & Fe & 15207.5 & 7.370 & 0.246 \\
J18571019+4848067 & Fe & 15224.7 & 7.130 & 0.090 \\
$\dotsb$&$\dotsb$&$\dotsb$&$\dotsb$&$\dotsb$\\
\hline
\hline 
\end{tabular}
\label{tab:online_lines}
\\ \\ \tablefoot{ This table contains the line-by-line abundance information for every stars in the APOKASC sample. This table is online-only. A portion is shown here to indicate form and content.}
\end{table}

Table~\ref{tab:onlinerec} contains a illustration of the recommended stellar parameters and chemical abundances for the APOKASC sample.  The \teff, \logg, and mass are taken from \cite{Pinsonneault2014}. The convolution term (which includes information about the \vsini, instrument broadening, and \vmac, see Sect. \ref{subsec:BACCHUS} for more details) and its uncertainty are also included in this table. The final abundances for C, N, O, Mg, Ca, Si, Ti, S, Al, Na, Ni, Mn, Fe, K, P, Cr, Co, Cu, and V as described in Sect. \ref{subsec:Chemistry} are provided with the formal uncertainties. The uncertainties provided in the table for the abundances are the standard error in the mean for elements which have more than 1 line and the method-to-method scatter in all other cases. To account for the full error budget, one should combine these in quadrature with the uncertainties in the abundance caused by uncertainties in the stellar parameters (a typical values for these can be found in Table~\ref{tab:abunderror}). Identical tables for cluster stars will also be electronically available. 


\begin{sidewaystable*}[ht]
  \setlength{\tabcolsep}{3pt}

\caption{Line Abundances Online Table Format.} 
\centering 
\begin{tabular}{c c c c c c c c c c c c c c c c c c    } 
\hline\hline 

APOGEE ID & KIC ID & Mass& \teff(K) & \logg & [Fe/H] & e[Fe/H] & convol & econvol & \vmic & e\vmic & [C/H] & e[C/H] & [N/H] & e[N/H] & [O/H] & e[O/H] & $\dotsb$\\
\hline

J18583782+4822494&10907196&1.500&4740&2.543&-0.26&0.11&13.546&0.049&1.720&0.007&-0.31&0.04&0.03&0.09&-0.16&0.06&$\dotsb$\\
J18582020+4824064&10962775&1.210&4736&2.444&-0.28&0.17&13.61&0.188&0.790&0.288&-0.31&0.09&-0.33&0.08&-0.49&0.20&$\dotsb$\\
J18571019+4848067&11177749&1.080&4644&2.427&-0.03&0.11&13.453&0.003&1.273&0.011&-0.28&0.13&0.15&0.08&0.13&0.20&$\dotsb$\\
J18584464+4857075&11231549&1.550&4541&2.371&-0.05&0.09&13.64&0.004&1.098&0.003&-0.12&0.09&0.12&0.10&0.04&0.13&$\dotsb$\\
J18582108+4901359&11284798&1.290&4283&1.854&0.22&0.14&13.265&0.505&0.622&0.015&0.12&0.21&0.41&0.12&0.44&0.09&$\dotsb$\\
J18583500+4906208&11337883&1.530&4802&3.076&-0.17&0.06&13.809&0.009&1.308&0.002&-0.13&0.04&-0.08&0.11&-0.03&0.08&$\dotsb$\\
J18590205+4853311&11178396&0.860&4824&2.359&-0.69&0.12&16.335&0.001&1.372&0.002&-0.55&0.09&-0.35&0.05&-0.33&0.13&$\dotsb$\\
J18581445+4901055&11284760&1.140&4658&2.433&-0.06&0.09&14.375&0.395&1.312&0.011&-0.12&0.17&-0.27&0.10&-0.21&0.06&$\dotsb$\\
J19010271+4837597&11072470&1.100&4558&2.403&0.16&0.16&13.095&0.054&1.251&0.007&0.17&0.16&0.28&0.11&0.28&0.08&$\dotsb$\\
J19004144+4836005&11072334&1.610&4716&2.794&0.05&0.10&13.283&0.072&1.277&0.015&-0.17&0.09&0.54&0.10&0.27&0.01&$\dotsb$\\
J19005385+4831394&11017831&1.020&4829&3.257&-0.17&0.10&12.464&0.021&1.300&0.008&-0.31&0.08&0.06&0.10&-0.08&0.09&$\dotsb$\\
J19005306+4856134&11232325&1.360&4588&2.672&0.14&0.11&13.126&0.141&1.195&0.032&-0.01&0.16&0.32&0.08&0.20&0.02&$\dotsb$\\
J19013400+4908307&11339000&1.330&4713&2.480&-0.03&0.10&13.588&0.033&1.171&0.036&-0.23&0.15&0.47&0.09&0.32&0.04&$\dotsb$\\
J19003958+4858122&11232225&1.300&4581&2.410&0.29&0.10&13.796&0.465&1.063&0.007&0.09&0.14&0.42&0.09&0.25&0.19&$\dotsb$\\
J19005590+4905481&11285650&1.250&4606&2.865&0.12&0.08&13.568&0.412&1.632&0.019&0.11&0.16&0.27&0.10&0.18&0.07&$\dotsb$\\
J19012632+4914106&11391750&1.260&4744&2.833&-0.19&0.08&13.354&0.017&1.029&0.015&-0.32&0.27&-0.03&0.08&-0.08&0.03&$\dotsb$\\
J19022554+4832549&11018481&1.240&4399&1.864&-0.02&0.12&13.003&0.003&0.955&0.001&-0.15&0.16&0.18&0.08&0.15&0.04&$\dotsb$\\
J19015877+4838074&11072852&1.410&4453&1.861&0.02&0.13&13.006&0.009&1.234&0.058&-0.27&0.14&0.45&0.06&0.19&0.22&$\dotsb$\\
J19032139+4847102&11126673&1.220&4602&2.630&-0.16&0.09&12.625&0.086&1.007&0.006&-0.03&0.08&0.21&0.10&-0.02&0.07&$\dotsb$\\
J19024750+4850053&11179815&1.250&4715&2.456&-0.15&0.08&13.081&0.181&1.253&0.005&-0.26&0.04&0.02&0.08&-0.02&0.05&$\dotsb$\\
J19025982+4834262&11018710&1.350&4561&2.618&0.12&0.10&12.592&0.359&1.070&0.01&0.05&0.04&0.40&0.08&0.19&0.19&$\dotsb$\\
J19032841+4845265&11126721&nan&4486&nan&nan&nan&nan&nan&nan&nan&nan&nan&nan&nan&nan&nan&$\dotsb$\\
J19041642+4850142&11180468&1.560&4617&2.598&0.30&0.12&12.912&0.132&0.821&0.063&-0.02&0.17&0.55&0.08&0.17&0.13&$\dotsb$\\
J19042303+4843556&11127105&1.030&4686&3.175&0.04&0.10&12.801&0.01&0.988&0.003&0.03&0.11&0.03&0.09&nan&nan&$\dotsb$\\
J19040509+4850191&11180378&0.960&4618&2.431&0.06&0.11&12.709&0.078&0.993&0.004&-0.03&0.01&0.18&0.07&0.13&0.02&$\dotsb$\\
J19053776+4843064&11127586&1.220&4782&2.795&-0.47&0.08&14.022&0.045&1.177&0.016&-0.55&0.13&-0.24&0.08&-0.25&0.11&$\dotsb$\\
J19042456+4907335&11340165&1.060&4725&2.417&-0.31&0.08&13.31&0.006&1.339&0.003&-0.34&0.02&-0.17&0.10&-0.14&0.05&$\dotsb$\\
J19052821+4848200&11180994&1.380&4617&2.730&0.01&0.09&13.635&0.016&1.252&0.007&-0.07&0.18&0.19&0.10&0.10&0.00&$\dotsb$\\
J19045132+4906064&11340377&0.910&4599&2.233&-0.30&0.09&12.991&0.04&1.041&0.073&-0.21&0.06&-0.20&0.12&-0.12&0.14&$\dotsb$\\
J19062780+4903417&11287844&1.680&4729&2.543&-0.11&0.09&13.564&0.34&0.968&0.009&-0.19&0.09&0.24&0.11&0.03&0.15&$\dotsb$\\
$\dotsb$&$\dotsb$&$\dotsb$&$\dotsb$&$\dotsb$&$\dotsb$&$\dotsb$&$\dotsb$&$\dotsb$&$\dotsb$&$\dotsb$&$\dotsb$&$\dotsb$&$\dotsb$&$\dotsb$&$\dotsb$&$\dotsb$&$\dotsb$\\
\hline
\hline 

\end{tabular}

\label{tab:onlinerec}
\tablefoot{This table contains the stellar parameters and abundance information for every star in the APOKASC sample. The \teff, \logg, and Mass are taken from \cite{Pinsonneault2014}. The convolution parameter (which combines the \vsini, instrument broadening and \vmac) and its uncertainty can be found in Columns 8 and 9, respectively. Only formal uncertainties are listed here. This table is online-only. A portion is shown here to indicate form and content.}
\end{sidewaystable*}

%

%

\end{appendix}

\begin{acknowledgements} 
K.H. is funded by the British Marshall Scholarship program and the King's College, Cambridge Studentship. This work was partly supported by the European Union FP7 programme through ERC grant number 320360. KH thanks A. Casey and D. Stello for very fruitful conversations and comments that helped clarify the manuscript. The research leading to the presented results has received funding from the European Research Council under the European Community?s Seventh Framework Programme (FP7/2007- 2013)/ERC grant agreement (No 338251, StellarAges).
\end{acknowledgements}

\bibliographystyle{aa} 
\bibliography{bibliography} 

\end{document}